\documentclass[letterpaper]{article} % DO NOT CHANGE THIS
\usepackage{aaai2026}  % DO NOT CHANGE THIS
\usepackage{times}  % DO NOT CHANGE THIS
\usepackage{helvet}  % DO NOT CHANGE THIS
\usepackage{courier}  % DO NOT CHANGE THIS
\usepackage[hyphens]{url}  % DO NOT CHANGE THIS
\usepackage{graphicx} % DO NOT CHANGE THIS
\usepackage{natbib}  % DO NOT CHANGE THIS AND DO NOT ADD ANY OPTIONS TO IT
\usepackage{caption} % DO NOT CHANGE THIS AND DO NOT ADD ANY OPTIONS TO IT
\usepackage{algorithm}
\usepackage{algorithmic}

\usepackage{newfloat}
\usepackage{listings}
\DeclareCaptionStyle{ruled}{labelfont=normalfont,labelsep=colon,strut=off} % DO NOT CHANGE THIS
\floatstyle{ruled}
\newfloat{listing}{tb}{lst}{}
\floatname{listing}{Listing}
\usepackage{booktabs}
\usepackage{subcaption}
\usepackage{color}
\usepackage{amsmath}
\usepackage{multirow}
\usepackage{array}
\usepackage{tabularray}
\usepackage{enumitem}
\usepackage{amssymb}
\usepackage{tabularx}
\usepackage{tcolorbox}
\usepackage{listings}

\usepackage{xcolor}
\newcommand{\answerYes}[1]{\textcolor{blue}{#1}} 
\newcommand{\answerNo}[1]{\textcolor{teal}{#1}} 
\newcommand{\answerNA}[1]{\textcolor{gray}{#1}}

\title{Does Geo-co-location Matter? A Case Study of Public Health Conversations during COVID-19}
\author {
    Paiheng Xu,
    Louiqa Raschid,
    Vanessa Frias-Martinez
}
\affiliations {
    University of Maryland, College Park\\
    \{paiheng,lraschid,vfrias\}@umd.edu
}
\begin{document}

\maketitle

\begin{abstract}
Social media platforms like Twitter (now X) have been pivotal in information dissemination and public engagement.
The objective of our research is to analyze the effect of localized engagement on social media conversations.
This study examines the impact of geographic co-location, as a proxy for localized engagement.
Our research is grounded in a COVID-19 dataset.
A key goal during the pandemic for public health experts was to encourage prosocial behavior that could impact local outcomes such as masking and social distancing.
Given the importance of local news and guidance during COVID-19, we analyze the effect of localized engagement, between public health experts (PHEs) and the public, on social media.
We analyze a Twitter Conversation dataset from January 2020 to November 2021, comprising over 19 K tweets from nearly five hundred PHEs, and 800 K replies from 350 K participants. 
We use a Poisson regression model to show that geo-co-location is indeed associated with higher engagement.
%% Our findings reveal that higher engagement is observed, in particular, in conversations on topics including masking, lockdowns, or education, or those originated by academic or medical professionals. 
Lexical features associated with positive sentiment, emotion and personal experiences were more common in geo-co-located conversations. 
The relationship between geo-co-location and engagement also varies by the PHE’s profession.
To complement our statistical analysis, we also applied a large language model (LLM)-based method to explore how linguistic and thematic content varied by geo-co-location, offering a complementary lens on discourse differences.
This research provides insights into how geographic co-location influences social media engagement and can inform strategies to improve public health messaging.
\end{abstract}


%% \input{sections/1_intro}
\section{Introduction}

Social media platforms such as Twitter (now X) play a crucial role in information dissemination and public engagement.
Our research aims to analyze the effect of localized engagement on social media conversations.  
We use geographic co-location (defined in the paper) as a proxy for localized engagement.
The COVID-19 pandemic presented a complex communication challenge:
public health experts had to provide vital guidance, in a scenario where the science was constantly being updated, and in the midst of a social media deluge of mis- and dis-information.
A key goal for public health experts (PHEs) was to encourage pro-social behavior that could impact In Real Life (IRL) outcomes ~\cite{fischer2020local, miles2022using}. 
Given the importance of local news and guidance during COVID-19, our research aims to analyze the effect of geo-co-located engagement on social media conversations.  
%% We use geographic co-location (defined in the paper) between the public health expert (PHE) and the public, as a proxy for localized engagement.
Prior research has shown that physical distance between social media users impacts the likelihood of online interactions, but not necessarily the content of those interactions. 
For example, the likelihood of interaction decreases with greater physical distance~\cite{%%liben2005geographic,
leskovec2008planetary%%newman2011structure
}. 
%% and Twitter users tend to retweet more from those co-located in the same country ~\cite{kulshrestha2012geographic,cuevas2014understanding}.
\citet{bozarth2023role} found that news sharing and user interactions occur more often within a state on Reddit, when compared to across states.
While there has been some recent research on social media engagement during COVID-19 %% \cite{rao2023pandemic,bojja2020early,gallagher2021sustained}, 
there is limited research on conversations involving PHEs and the public \cite{rao2025public}. 

In this paper, we hypothesize that geographic co-location between public health experts (PHEs) and the public is associated with greater engagement on social media.
Our novel research advances the state of the art by analyzing the impact of geographic co-location between PHEs and the public on the following: 
(1) Engagement as measured by the count of participants and replies per conversation around an original tweet. 
(2) The linguistic style, emotional tone, and COVID-19 topics expressed in conversations.
(3) The profile (profession) of the public health expert. 
We note that our research is the first to study both the relationship between geographic co-location and engagement, and the content characteristics of conversations influenced by such geo-co-location.
% We note that our research is the first to study the role of geographic co-location on engagement.
The study also contributes to the literature on the effectiveness of public health messaging during COVID-19.

Our research is based on an X (Twitter) dataset spanning January 2020 through November 2021.
Starting with a seed set of thirty public health experts (PHEs), we identified a larger group of almost five hundred PHEs or their expert peers in adjacent domains, e.g., medical communication. 
We then curated a dataset of conversations, comprising over 19 K posts (tweets) from the PHEs, and approximately 800 K replies from almost 350 K participants in the conversations.
We used Carmen \cite{dredze2013carmen,zhang2022changes} to determine geo-location (US state).
We then identified geo-co-located pairs - where the PHE and participant are in the same US state,
and non-geo-co-located pairs, respectively. Linguistic Inquiry and Word Count (LIWC)~\cite{pennebaker2015development} was used to identify lexical features that capture levels of self-disclosure and psychological processes.
An LLM-based method \cite{zhong2022describing,zhong2023goal} was used to further explore how linguistic and thematic content varied by geo-co-location, 
offering a complementary lens on discourse differences.
%% Machine learning models ~\cite{rao2023pandemic} and LLMs were used to identify COVID-19 topics in the original post (tweet), and the profile (professions) of PHEs. 
Our analysis reveals the following key findings:
\begin{enumerate}[leftmargin=*]
    \item We find that engagement is significantly higher when the PHEs and the public are geo-co-located, compared to when they are not. 
    This result is based on a Poisson regression model predicting reply counts for each original tweet.
    The finding is robust when we apply a mixed-effects specification that accounts for author-level heterogeneity, and is also supported by a non-parametric statistical test comparing engagement rates across all tweets.
    % We do so using a Poisson regression model that predicts the count of replies for each original tweet and a non-parametric statistical test that compares the engagement rate across all tweets.
    % \item Conversations discussing masking, lockdown, and education, that may have a local impact, exhibit higher geo-co-located engagement, while vaccine conversations elicit more non-geo-co-located engagement.
    % \ph{Adjust based on regression results?}
    \item The sharing of personal experiences and emotional expressions, e.g., anxiety, in the original post, are associated with higher geo-co-located engagement. These findings are confirmed using a large language model (LLM) based approach to generate and validate hypotheses.
    \item Replies from geo-co-located participants are more positive in sentiment, and express more emotions and personal issues, when the PHEs initially share personal experiences / feelings.
    %% These findings are confirmed using an LLM based approach to generate and validate hypotheses.
	% \item Geo-co-location has a greater impact on engagement for PHEs in academia or medical professions, in comparison to PHEs in media and policy / political professions.
    \item The relationship between geo-co-location and engagement varies by the PHE's profession. 
    % After controlling for author-level heterogeneity, we find a positive association for political and academic professionals, a modest negative association for media professionals, and a non-significant association for medical professionals. 
\end{enumerate}

To summarize, this research provides novel insights on the impact of geo-co-location, on engagement. It has the potential to inform how public health messaging strategies can harness geo-co-location to enhance engagement and to maintain a positive discourse around public health guidance.

\section{Related Work}

\paragraph{Geo-co-location in social networks}
Many studies have shown that users on social media platforms tend to connect or interact preferentially with users who are geographically closer over those who are farther away~\cite{leskovec2008planetary,%%crandall2010inferring,scellato2011socio,
laniado2018impact}.
% Similar findings have also been replicated for users on cellphone social networks~\cite{hong2016topic,frias2012computing}.
Our study focuses on message-exchange interactions such as replying to an original post, on platforms such as Twitter, where interactions typically reflect a shared interest in specific topics. %%~\cite{kwak2010twitter}.
We specifically investigate geo-location of users; this differs from prior research on the correlation between user geo-location and content ~\cite{%%hu2015predicting,
pavalanathan2015confounds,cheke2020understanding}. 
Twitter interactions in the form of replies and mentions have been successfully used as predictors of geographically closer connections,%%~\cite{sadilek2012finding,mcgee2013location,jurgens2015geolocation}, 
but there is limited research on the influence of geo-co-location on user interactions. 
%% \citet{kulshrestha2012geographic} analyze the information flow between different countries by counting the number of information producers and consumers in each country and they consider all followers of a user as the consumers of the tweets shared by this user. Similarly, 
\citet{cuevas2014understanding} count the number of retweets and compare how tweets are shared across countries. 
%These studies are limited to a country-level analysis. 
\citet{bozarth2023role} studied the geographical diffusion of news articles on Reddit. %% is rare among states in the U.S. 
None of these studies investigated the specifics of social media interactions.
The data collection for this research was facilitated when Twitter released API v2 in late 2020; the API supported the retrieval of a conversation, i.e., retweets and replies to an original post.

\paragraph{Geo-location data on Twitter}
Twitter provides multiple sources of geo-location information in the metadata of tweets and users, including coordinates, a Place object within a tweet, and location information within a user profile.
Further, the geo-location of a tweet can also be inferred from its content~\cite{%%mahmud2012tweet,
%han2014text,rahimi2016pigeo,
izbicki2019geolocating}.
Given that our study focuses on the geo-location of users, we rely on tools that extract geo-location from user-provided metadata in tweets and user profiles~\cite{dredze2013carmen,zhang2022changes}.
We acknowledge that Twitter users
% ~\cite{wojcik2019sizing} 
and those who use Twitter's geo-location services %%~\cite{sloan2015tweets, pavalanathan2015confounds,karami2021analysis} 
may be a biased sample of the user population.
Despite that limitation, Twitter's geo-location data has been shown to be valuable in many applications such as disaster response~\cite{%%crooks2013earthquake,
hong2020modeling}, 
urban planning ~\cite{frias2012characterizing}
%milusheva2021applying
and public health surveillance~\cite{jordan2018using,xu2020twitter,xu2024twitter}.
%sigalo2023using,sigalo2023uptake}.

% \paragraph{Engagement on social media} \ph{This topic might be too broad. Ignoring it for now.}

% \paragraph{Content understanding for interactions on social media} 
% Numerous text analysis approaches including topic detection, linguistic analysis and text classification have been
% employed to comprehend user interactions on social media.
% %% topic analysis~\cite{
% % dahal2019topic,choi2023analyzing
% %% cheke2020understanding,li2024improving}, 
% %% linguistic analysis~\cite{pavalanathan2015confounds,choi2021more}
% % ,wood2021using,peterson2024linguistic
% %% , and text classification models for constructs like sentiment~\cite{ % burke2016once,lwin2020global}. 
% % and emotion~\cite{wheaton2021fear}. 
% Large Language Models (LLMs) have also been used for social media analysis ~\cite{ziems2024can}.
% %% In this work, we utilize a wide range of text analysis tools, including topic modeling~\cite{eisenstein2011sparse}, 
% In our research we use LIWC for linguistic analysis~\cite{pennebaker2015development}, 
% sentiment classification~\cite{loureiro-etal-2022-timelms}, 
% and LLM-based methods ~\cite{zhong2022describing,zhong2023goal}.

\paragraph{Content understanding of social media interactions} 
Numerous text analysis tools have been employed to comprehend user interactions on social media. This includes topic modeling ~\cite{
% dahal2019topic,choi2023analyzing
cheke2020understanding,li2024improving,li2025large}, linguistic analysis~\cite{pavalanathan2015confounds,wood2021using,choi2021more},
% peterson2024linguistic
text classification models for constructs like sentiment~\cite{
% burke2016once,
lwin2020global}.
% and emotion~\cite{wheaton2021fear}. 
More recently, large language model (LLM) based approaches have shown the potential for more nuanced and deeper analysis of social interactions ~\cite{ziems2024can,zhou2025emojis}.
In this research, we use a wide range of text analysis tools, including topic modeling~\cite{eisenstein2011sparse}, LIWC for linguistic analysis~\cite{pennebaker2015development}, sentiment classification~\cite{loureiro-etal-2022-timelms}, and LLM-based methods for hypothesis generation, in the context of studying geo-co-located engagement.
\section{Data}
\label{sec:data}
% Introduce the data collection process, definitions for related concepts, and basic data statistics.

The Twitter conversation dataset used in this study is centered around COVID-19 discussions.
The creation of the dataset commenced with a curated seed user list of 30 Public Health Experts (PHEs) (see Appendix \ref{app:phe}), both academics and health professionals, who were active in COVID-19 discussions.
These 30 experts were handpicked by academic colleagues in public health. 
PHEs include individuals with advanced degrees in medicine, epidemiology, genomics, infectious diseases, public policy, and economics~\cite{rao2025public}.
A retweet network of these 30 seed experts was then constructed using a dataset of over 1 billion publicly available COVID-19 Tweets \cite{chen2020tracking}. 
All users in the retweet network were ranked using eigenvector centrality \cite{ghosh2010predicting}, and the Top 500 ranked users were identified.
Eigenvector centrality measures a user’s influence based not only on how many times they are retweeted, but also on the influence of the users who are retweeting them.
We then manually removed organizations or bots.
We obtained an expanded set of 489 PHEs or PHE adjacent influential users.
An example of a PHE-adjacent user may be a health reporter, a policy expert, or an appointee in a health agency, as well as other influential users (high-centrality nodes) in the retweet network.
% The details can be found in \citet{}.

The next step was to collect all of the original posts (tweets) of PHEs, and the resulting conversations, comprised of replies\footnote{https://developer.twitter.com/en/docs/twitter-api/conversation-id}. The dataset covered a period from the early days of the pandemic, starting in January 21, 2020, through November 4, 2021.
We only consider the original posts (tweets) of the 489 PHEs, i.e., we exclude their own retweets, replies, and quoted tweets, giving a dataset of 144 K tweets.
The final step was to collect all of the replies for the original tweets.
Unfortunately, restrictions and rate limits imposed by Twitter on the academic API limited our ability to rehydrate all of the replies (conversations).
We were able to obtain all reply tweets for 19.5 K original tweets, from 462 PHEs, that had at least one reply.
The final dataset included approximately 786 K reply tweets from 345 K unique participants in conversations with the 462 PHEs. 
Each tweet received an average of 40.24 replies, with a median of 5 replies per original PHE tweet, and approximately 70\% of the tweets did not have replies.
%% The distribution of replies ranged widely, from a minimum of 1 reply to a maximum of 11.7 K replies.

In this paper, we refer to an original PHE tweet as a {\it tweet} and a reply tweet as a {\it reply}. 
We refer to the PHEs as either {\it authors} or {\it PHE authors}. 
All users who responded with a reply tweet to an original tweet are referred to as {\it participants} in the conversation.
%DONE: \lr{Use tweet not Tweet? Use participant not replier?}
Finally, we refer to the union of participants who reply to any of the original tweets of a PHE author as the {\it audience} of the PHE author.

% \section{Measuring the Impact of Geo-co-location on Engagement}
%% \section{Engagement Rate}
\section{Geo-Co-Location and Engagement}
%% \subsection{Definitions}

We first compare the activity statistics of PHE authors and participants who disclose geo-location and those who do not. 
We then define the geo-co-located (gcl) audience and participant count (reply count) for a tweet. 
We use a Poisson regression model that predicts the reply count to demonstrate the higher engagement of gcl participants.
We then define the engagement rate and use the Kolmogorov-Smirnov (KS) test ~\cite{hodges1958significance} to determine the
statistical difference between the gcl and non-gcl engagement rates. 
% Prior research has shown that social media users who share their location information are more active~ \cite{rzeszewski2017spatial}.

\subsection{Geo-Located Activity Statistics}

We use the Carmen library \cite{dredze2013carmen, zhang2022changes} to infer the geo-location of the PHE authors and the participants.
Carmen uses the following features: location information and self-description from the user profile, coordinates, and the Place object from the metadata associated with tweets.
The most frequently mentioned (inferred) state is selected as the home state for the user.
We note that Carmen could infer the geo-location (state) of over $74\%$ of the PHEs (Table~\ref{tab:stats}) and over $1/3$ of the participants (Table~\ref{tab:stats2}).
We chose to conduct our analysis at the state level to ensure data reliability. The accuracy of the Carmen tool is substantially higher at this level (over $99\%$) compared to city-level analysis (around $53\%$).
% Carmen reports that user geolocation accuracy is over $99\%$ at the granularity of the US state when geo-related metadata is available in the user metadata.

%\textcolor{red}{To me, the next two paragraphs would go in the Results section. If we are showing that participants with state information post more replies, that is a preliminary analysis to answer our research question, I would say. I think I would keep table 1 because it is just general statisics, but I would move the statistical discussion to Results. }

%\lr{Paiheng, Can you switch the two following paragraphs? First present number of replies per participant? If you do that remember to move the KS test explanation as well.}

We report on activity statistics comparing PHE authors / participants who share / do not share geo-location, in Tables \ref{tab:stats} and \ref{tab:stats2}, respectively. 
The mean and standard deviation (std) of the count of replies per participant,
%Of greater interest, however, is that the average number of replies, per participant, 
for tweets from PHE authors with geo-location, is $2.29 \pm 6.55$, in comparison to $1.91 \pm 7.82$ for tweets from PHE authors without geo-location. 
We used the two-sample Kolmogorov-Smirnov (KS) test ~\cite{hodges1958significance},
a non-parametric statistical test, to assess the statistical significance of the difference
between the samples. 
It rejected the null hypothesis ($p<0.001$), pointing to a statistically significant increase in engagement from participants for PHE authors with geo-location.
%observed between the replies distributions for geo-located and non-geo-located authors 
%significantly higher, in comparison to tweets from authors without geo-location ($2.29$ vs. $1.91$). The corresponding two-sample KS results rejected the null hypothesis ($p<0.001$), indicating that geo-location information from authors increases participant activities.
The mean and std of the count of replies per tweet,
for PHE authors with geo-location is $23.08 \pm 53.93$, in comparison to $25.98 \pm 60.79$ for PHE authors without geo-location.
However, that difference is not statistically significant.
% (the null hypothesis was not rejected with p-value $>$ 0.01).
We conclude that geo-location information for PHE authors does appear to impact engagement at the participant level, with higher engagement for PHEs that have geo-location information.
% however, that impact is not significant when measuring the intensity of responses. 

%Our first hypothesis is to determine if these two distributions of replies are drawn from the same %population.
%We use the two-sample Kolmogorov-Smirnov (KS) test~\cite{hodges1958significance},
%%DONE:\textcolor{red}{add REF}
%a non-parametric statistical test, to test the null hypothesis that the mean of replies for both groups of %authors come from populations with the same distribution. 
%The test did not reject the null hypothesis (p-value $>$ 0.01), finding no significant difference in the %average number of replies.
%%whether these two groups of tweets come from the same distribution in terms of receiving replies and the p-%value is not significant.
%A possible explanation is that the PHEs in the COVID-19 dataset are both active and influential; as a %result, geo-location information does not have a significant impact on attracting engagement (replies). 
%\ph{The following is an update of a previous analysis. Please review if it is clear.}

%DONE: \textcolor{red}{Paiheng, the terms in Table 1 and 2 are important, so I would not abbreviate them. Maybe try to break the column names into two lines? The columns should say "Mean (Std) replies per Tweet and Mean (Std) replies per Participant. and in table 2 change Part. to Participants. also, it's either all column names with capital letters or none :)}

\begin{table*}[!htb]
\centering
% \small
\resizebox{\linewidth}{!}{%
\begin{tabular}{lrrrrrrr} \toprule
~          & \# Authors & \# Tweets & \# Replies & \# Participants & Mean (std) followers & Mean (std) replies per tweet & Mean (std) replies per participant \\ \midrule
With state & 342        & 16,465            & 380k            & 166k            & 157k (654K)         & 23.08 (53.93)   & {\bf 2.29 (6.55)}       \\
No state   & 120        & 4,432             & 115k            & 60k             & 160k (438K)         &  25.98 (60.79)  & {\bf 1.91 (7.82)}    \\ \bottomrule
\end{tabular}
}
\caption{Statistics for PHEs with and without geo-location in the COVID-19 Conversation Dataset. We {\bf bold} the cases where the comparison between the PHEs with and without state is significant using the two-sample KS test ($p<0.001$).}
\label{tab:stats}
\end{table*}

\begin{table}[!htb]
\centering
\small
\begin{tabular}{lrr} \toprule
           & \# Participants & Mean (Std) replies  \\ \midrule
With state & 73,912                               & {\bf 2.45 (8.33)}\\
No state   & 131,760                              & {\bf 2.38 (7.43)}\\\bottomrule
\end{tabular}
\caption{Statistics for participants with and without geo-location in the COVID-19 Conversation Dataset.  We {\bf bold} the cases where the comparison of participants with and without state is significant using the two-sample KS test ($p < 0.05$).}
\label{tab:stats2}
\end{table}

%DONE: \lr{Paiheng be consistent in showing mean only or meam and std dev. You are not consistent!}

Next, we consider the statistics for the {\it participants} in Table \ref{tab:stats2}.
We observe that participants with geo-location information post significantly more replies,
with a mean and standard deviation of $2.45 \pm 8.33$,
than participants without geo-location, with  mean and std of 
%% $2.45 \pm 8.33$ vs. 
$2.38 \pm 7.43$.
This difference is statistically significant (two-sample KS test rejected the null hypothesis with $p<0.05$).
%, $p<0.001$ based on a KS test. 
% Geo-located participants have a higher number of replies (engagement).

% \ph{More on geo-location average. Repeated/active participants have geo-location.}
%DONE: \lr{Good idea to add more evidence about the importance of sharing location.}

% \begin{figure}[h]
%     \centering
%     \includegraphics[width=0.65\linewidth]{figures/participants_replies.png}
%     \caption{Number of replies for geolocated participants vs. non-geolocated participants.}
%     \label{fig:enter-label}
% \end{figure}

To summarize, these activity statistics reflect that sharing geo-location information is an important signal for engagement in social media conversations, for both PHE authors and participants.
This finding is also consistent with prior results that social media users who share their location information are more active ~\cite{rzeszewski2017spatial}.

\subsection{A Model of Engagement}
\label{sec:engagement}

\paragraph{Geo-co-location (\textit{gcl})} 
For each (tweet, reply) pair in the dataset, we refer to the participant being geo-co-located (\textit{gcl}) with the author, if they are both located in the same state. 
Conversely, the participant and the author are not geo-co-located (\textit{non-gcl}) if they are located in different states. 
Participants without geo-location are excluded from the \textit{gcl} and \textit{non-gcl} groups for further analysis.

\paragraph{Audience}
For a tweet $T$ from author $A$, we define the \textit{gcl audience} $W$ as the set of all \textit{gcl participants} across all the tweets created by $A$ in the dataset. 
We can similarly define the \textit{non-gcl audience} for $A$.
Since the popularity of an author can vary over time, we define the audience as a cumulative measure over the time period of the data collection.
We note that while {\it gcl} participants are limited to those from the same state as the tweet author, {\it non-gcl} participants can be from all other states.
Hence, the {\it gcl} audiences are inherently smaller than the {\it non-gcl} audiences.

% \textcolor{red}{
We model engagement using a Poisson regression model that can predict the count of \textit{gcl} and \textit{non-gcl} participants (reply count) while controlling for the difference in the audience count.
For each tweet, we treat the count of {\it gcl} and {\it non-gcl} replies as separate data points for the regression model. 
Each data point is also associated with a categorical variable indicating the origin as the {\it gcl} or {\it non-gcl} audience. 
\footnote{The model assumes that the decision to reply from the two audiences of each author are independent. Given that more than $40\%$ of replies in the dataset are directly replying to the original tweets and $95\%$ of the participants are {\it non-gcl} or {\it no-loc}, interactions between the {\it gcl} and {\it non-gcl} audience are expected to account for only a small proportion of the replies.}
Thus, the dependent variable is the count of {\it gcl} or {\it non-gcl} participants (replies), denoted as $Y_T$, and the independent variable, $X_T$, is the audience type ({\it gcl = 0} and {\it non-gcl = 1}).
To control for the size of the audiences, we include the corresponding audience count $W_T$ as an exposure variable. The model is formulated as follows:
\begin{equation}
\label{eq:poisson}
    \log(\mathbb{E}[Y_T]) = \beta_0 + \beta_1 \cdot X_T + \log(W_T).
\end{equation}
We can interpret $e^{\beta_1}$ as the multiplicative difference in the expected engagement between the two groups, accounting for exposure.

We also fit a mixed-effects Poisson regression \cite{seabold2010statsmodels} to account for author heterogeneity, which extends Eq. (\ref{eq:poisson}) by including a random intercept ($u_j$) for each author:
\begin{equation}
\label{eq:poisson_mixed}
\log(\mathbb{E}[Y_{ij}]) = \beta_0 + \beta_1X_{ij} + u_j + \log(W_{ij}),
\end{equation}
where the variables correspond to those in the main model but are indexed by tweet $i$ and author $j$.

The regression results in Table \ref{tab:poisson-results} (simpler model) show that the engagement of the {\it non-gcl} audience is $e^{-0.12} \approx 88.6\%$ of that of {\it gcl} audience, 
while controlling for the audience size ($N=31,852$, $p<0.001$).
\footnote{
We also fitted a Negative Binomial model with the same setup and obtained similar results. However, the estimated dispersion parameter 
$\alpha$ was very small, indicating minimal over-dispersion, suggesting that the Poisson model is more appropriate.
}
For example, if the engagement (replies per original tweet) for the {\it gcl} group is $x$, then the engagement for the {\it non-gcl} group would be approximately $.886 \times x$.
This indicates that the {\it non-gcl} audience tends to engage at a lower rate compared to the {\it gcl} audience, even after accounting for differences in the audience size.
The mixed-effects model further confirms our findings. See Appendix \ref{app:poisson} for details.

\begin{table}[!htb]
\centering
\resizebox{\linewidth}{!}{
\begin{tabular}{lclc} \toprule
Variable & Coefficient & IRR & 95\% CI (IRR)         \\ \midrule
Intercept       & -4.92       & 0.007       & (0.007, 0.007)     \\
$X_T (\textit{gcl} = 0, \textit{non-gcl} = 1)$     & -0.12      & {\bf 0.886}       & (0.870, 0.901)     \\ \bottomrule
\end{tabular}
}
\caption{Poisson regression model to predict the effect of engagement type (gcl=0; non-gcl=1) on the count of replies per original tweet. IRR is the incidence rate ratio. Statistically significant results ($p < 0.001$) are shown in bold.}
\label{tab:poisson-results}

\end{table}

\subsection{Comparison of Engagement Rates}
\label{sec:gcl_vs_non-gcl}

The Poisson regression model demonstrated the greater engagement of the gcl audience across all tweets in the dataset. 
We are further interested in comparing the difference between gcl and non-gcl engagement, per original tweet, 
so as to identify the characteristics of interesting subsets that exhibit the greatest difference in the level of engagement.
We use a measure of engagement rate for this analysis.

\paragraph{Engagement Rate}
The \textit{gcl engagement rate} for a tweet $T$ from author $A$ is defined as 
the fraction of $A$'s \textit{gcl audience} that replied to the tweet $T$:
$$
\textit{gcl engagement rate} = \frac{\text{count of } \textit{gcl participants } \text{for } T}{\text{count of } \textit{gcl audience } \text{for } A}
$$
\noindent 
The \textit{non-gcl engagement rate} for tweet $T$ of author $A$ is similarly defined using the \textit{non-gcl} participants and \textit{non-gcl} audience.
%%\textcolor{red}{The metric captures what fraction of the author's total {\it gcl} (or {\it non-gcl}) audience engaged with the specific tweet $T$.}

% \paragraph{Justification for Engagement Rate}
% \label{sec:justification}
% \textcolor{red}{review this. The reason why we chose the more complicated definition was to be able to compare gcl and non-gcl behaviors. }
The {\it gcl} (or {\it non-gcl}) engagement rate for tweet $T$, representing the fraction of audience participation, enables a direct comparison between the two levels of engagement per tweet.
A direct comparison would not have been possible using the un-normalized \textit{gcl} and \textit{non-gcl} reply counts from the previous Poisson model. 
On average, the \textit{gcl} replies take up 
$5.7\%$ 
of the replies for a tweet while \textit{non-gcl} replies take up 
$29.4\%$ .
For this computation, we filter the dataset to only include authors that have both audience types,
which ensures that for every tweet in the analysis, both {\it gc}l and {\it non-gcl} engagement rate can be calculated.
The filtered subset includes about $97\%$ of the tweets from authors with state information.

Table~\ref{tab:ER} displays the engagement rates (ER) for \textit{gcl} and \textit{non-gcl} audiences, as well as for audiences where the location of the participant could not be inferred (labeled as \textit{no-loc}).
The table shows that a higher average engagement rate 
%(higher \% of the potential audience) 
is observed for \textit{gcl} audiences, in comparison to \textit{non-gcl} audiences (mean values: 1.89\% vs 1.76\%).
%\lr{Paiheng the table has 1.76?}
A two-sample KS test confirmed the significance of this difference ($n=15,926$, $p<0.001$). 
%Table~\ref{tab:ER} also reports on the engagement rate of the potential audience where the location of the participant could not be inferred; it is labeled as \textit{no-loc}. 
In addition, the \textit{no-loc} engagement rate is very similar to the \textit{non-gcl} rate (mean values: 1.76\% vs 1.78\%).
This validates that our focus on comparing \textit{gcl} versus \textit{non-gcl} subsets is a reasonable experimental filter.

\begin{figure}[!htb]
    \centering
    \includegraphics[width=\linewidth]{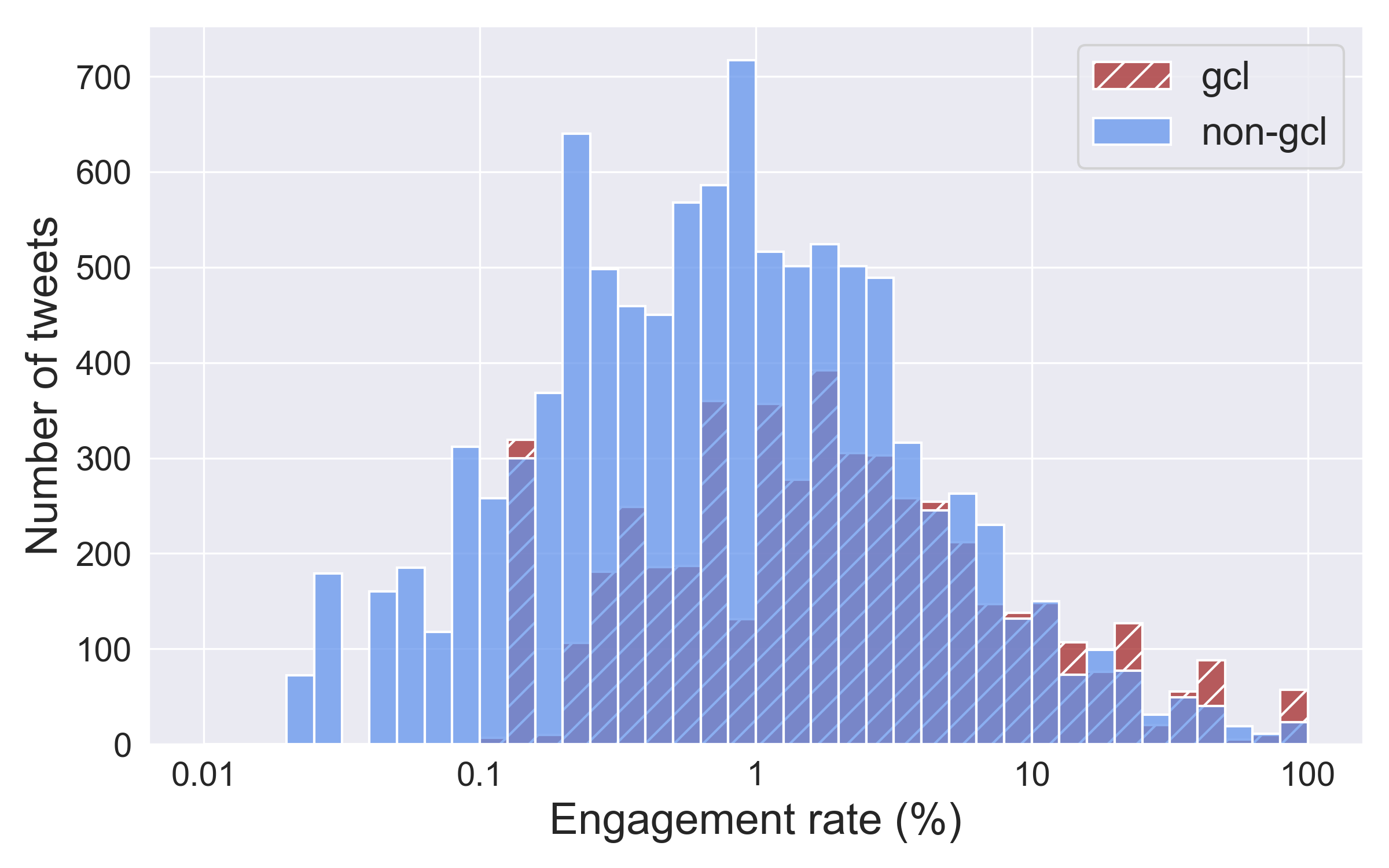}
    \caption{Distribution of \textit{gcl} and \textit{non-gcl} engagement rate. Tweets with $0\%$ engagement rate are not shown in the plot.}
    \label{fig:dist_engage}
\end{figure}

Figure~\ref{fig:dist_engage} provides a histogram of the \textit{gcl} (red hatched bars) and \textit{non-gcl} (blue bars) engagement rates.
%DONE: \lr{IMPORTANT Paiheng change the label / caption in the figure to include gcl and non-gcl!!}
%DONE: \lr{The X axis is strange? Explain what -8 and 0 represent in terms of engagement rate? Include the explanation in the text and the caption for the figure.}
The significant rightward skew of the \textit{gcl} histogram clearly illustrates its dominance in the region with a higher engagement rate region; for example, we see the histogram extend to a 100\% engagement rate. 
%DONE: \textcolor{red}{I would explain the figure a little bit more. LR: Good!}

In summary, the Poisson regression model of participant counts, the statistical significance of the \textit{gcl} versus \textit{non-gcl} engagement rate distributions using the KS test, and the insights from the histograms, confirm our hypothesis that geographical co-location between PHE authors and participants is associated with greater engagement on social media. 

\begin{table}[!htb]
\footnotesize
\centering
\resizebox{\linewidth}{!}{%
\begin{tabular}{lccr} \toprule
~              & Mean (Std) participants per tweet  & Mean (Std) ER  \\ \midrule
{\it gcl}    & 0.99 (3.32)           & {\bf 1.89\% (8.05\%)}                                   \\
{\it non-gcl} & 5.16 (13.09)         & {\bf 1.76\% (6.10\%)}                                  \\
{\it no-loc}    & 11.18 (26.56)      & 1.78\% (5.53\%)                                   \\ \bottomrule
\end{tabular}
}
\caption{Engagement rate (ER) for participants. \textit{no-loc} means there is no location information for the participants. We {\bf bold} the statistic comparison between \textit{gcl} and \textit{non-gcl} groups that is significant from a two-sample KS test ($p < 0.001$).}
\label{tab:ER}
\end{table}

\section{The Nuances of Geo-co-location on Engagement}

Having established that there is higher engagement of the \textit{gcl} audience, we next explore the nuances of this engagement as follows:
First, we explore whether greater \textit{gcl} (\textit{non-gcl}) engagement is dependent on the type of COVID-19 issue being discussed.
We then drill down into the content, to identify lexical features that may reveal additional insights into the emotions and concerns that are expressed in \textit{gcl} versus \textit{non-gcl} conversations.
Finally, we determine if the PHE author profile, and in particular, their profession, has an impact on attracting greater levels of \textit{gcl} engagement.

\subsection{What Tweets attract more \textit{gcl} engagement?}
\label{sec:tweet}

\paragraph{Impact of COVID-19 Topics}

We first examine whether the \textit{gcl} engagement rates vary across Tweets with different topics, and how these variations relate to \textit{gcl} versus \textit{non-gcl} engagement. 
%% Given that the conversation dataset is centered around COVID-19,
We adopt a method from \citet{rao2023pandemic} to identify tweets that discuss different COVID-19 topics. Their method extracts issue-relevant keywords from Wikipedia articles by identifying the most distinctive keywords when comparing issue-specific articles with general articles on the pandemic and politics in the U.S. using SAGE~\cite{eisenstein2011sparse}. These manually verified keywords are then used to label tweets through keyword matching. \citet{rao2023pandemic} validate their method by evaluating its results on a test sample with manual annotations.
% DONE: \textcolor{red}{in covid-19? or in other context?} 
We replicate this method and retain topics that have F1 scores higher than $0.7$;
the topics are \textit{masking}, \textit{lockdowns}, \textit{education}, and \textit{vaccines}. 
Table \ref{tab:sample_issue} shows sample tweets.
%DONE: I would add a couple of sentences explaining better the keyword matching method because I did not understand where the F1 score is coming from.}

% \lr{I suggest that the rest of this subsection should be RED!}

% We report on the \textit{gcl} and \textit{non-gcl} engagement rates on each of the subsets of tweets for these four topics in Table ~\ref{tab:er_issues}. We observe that tweets on masking have the highest engagement rates overall, and present the largest gap between these two rates.
A descriptive comparison of engagement rates for tweets discussing the four topics is shown in Table~\ref{tab:er_issues}.
We observe that tweets on masking are associated with the largest gap between \textit{gcl} and \textit{non-gcl} engagement.
This is consistent with \citet{rao2023pandemic} that finds that masking-related tweets gained more replies. 
Similarly, lockdown and education also show higher engagement rates for \textit{gcl} audiences.
On the other hand, {\bf vaccine-related tweets are associated with more engagement from non-gcl audiences}.
These differences were all statistically significant through bivariate KS tests.   
The discussions around topics such as masking, lockdowns, and education often had a strong local angle, and this may have led to a greater level of \textit{gcl} engagement. In contrast, vaccine-related discussions often take place within the context of a national policy \cite{karami2021covid}.

To test the robustness of these associations, we employ a multivariate linear regression model to examine the relationship between topics and engagement while controlling for an author's number of followers, a recognized driving factor in social media engagement \cite{cha2010measuring,bakshy2011everyone}.
We define the model as:
\begin{equation*}
    \text{Engagement Rate} = \beta_0 + \sum_{T_i \in \text{Topics}}(\beta_{i}T_i) + \beta_cF + \epsilon,
\end{equation*}
where $T_i$ is the binary indicator for each topic, $F$ is a control for the author's min-max-normalized follower,and $\beta_0$ and $\epsilon$ are the constant and error term respectively.
We repeat the analysis for three types of engagement rate as the dependent variable, i.e., \textit{gcl}, \textit{non-gcl} and the difference between \textit{gcl} and \textit{non-gcl} (denoted as {\it Diff}). 
Table \ref{tab:topic_regression} (Appendix \ref{app:full_liwc}) shows that, after controlling for authors' followers, only the topic of vaccines remains a statistically significant factor.\footnote{The results are consistent with the topic coefficients shown in our full lexical model in Table~\ref{tab:liwc_regression}.} This suggests the association between vaccine discussions and \textit{non-gcl} audience is a more robust finding, while the initial associations observed for the other topics are not significant when author popularity is controlled in the model.

\begin{table}[!htb]
\resizebox{\columnwidth}{!}{
\centering
\begin{tabular}{lrrrr} \toprule
             & Masking & Lockdowns & Education & Vaccines  \\ \midrule
\textit{gcl} ER     & 2.32\% & 1.49\%    & 1.94\%    & 1.52\%             \\
\textit{non-gcl} ER & 1.88\%  & 1.39\%    & 1.80\%    & 1.73\%             \\
{\bf \textit{Diff}}       & {\bf 0.44\%}  & {\bf 0.10\%}    & {\bf 0.14\%}    & \textbf{-0.21\%} \\
\# authors & 162     & 107       & 168       & 202                \\
\# Tweets  & 929     & 326       & 812       & 2240               \\ \bottomrule
%{\it Diff}       & 0.44\%  & 0.10\%    & 0.14\%    & \textbf{-0.21\%}   \\ \bottomrule
\end{tabular}
}
\caption{Difference between \textit{gcl} and \textit{non-gcl} engagement rate (ER) for Tweets discussing different COVID-19 topics. Significant ($p<0.001$) KS test results for \textit{gcl} ER and \textit{non-gcl} ER are bolded.}
\label{tab:er_issues}
\end{table}

\paragraph{Impact of Lexical Features} 
% \ph{Repeat on Non-GCL; Add control: followers; LIWC features from replies; mixed effect regression (group by authors)} 
Our next step is to drill down into the content, to identify lexical features that may reveal additional insights into the emotions and concerns expressed in \textit{gcl} versus \textit{non-gcl} conversations.
LIWC has been extensively used to study psychological patterns and emotional states in social media texts, providing insights into user behavior and social dynamics~\cite{schwartz2013personality,jiang2018linguistic,choi2021more}.
To explore whether psychological and emotional states - measured via LIWC categories - might affect the gcl engagement rates, when compared to non-gcl audiences, we propose the following approach. 
%how other aspects of language usage might be related to \textit{gcl} engagement, 
We count the occurrences of the LIWC categories in each original Tweet and normalize them by the number of tokens in the Tweet. 
We use the TweetTokenizer from NLTK \cite{bird2009natural} for tokenization.
Next, we run coefficient regression analyses with the occurrences of LIWC categories in the original Tweets as independent variables, and engagement rates as dependent variables, to assess the effect of LIWC categories on engagement rates. 

% \textcolor{red}{
We determine a list of LIWC categories relevant to the COVID-19 context based on a preliminary lexical analysis and existing literature.
Categories in LIWC \cite{pennebaker2015development} encompass two dimensions: grammar and psychological processes. We select several subcategories in grammar and most of the subcategories in psychological processes. 
% }
We include: first person singular pronouns (\texttt{i}),  first person plural pronouns (\texttt{we}), and second person pronouns (\texttt{you}) because the usage of pronouns reveal levels of self-disclosure~\cite{de2014mental,wang2016modeling} and they are also some of the most distinctive features in the tweets with high \textit{gcl} engagement rate based on our preliminary analysis (see Appendix \ref{app:lexical}). 
Additionally, we choose subcategories in psychological processes potentially relevant to COVID-19 discussions. \footnote{
We exclude subcategories that are hard to interpret in this context, such as gender references, cognitive processes, time orientations, and relativity.}
These include: (1) all subcategories in the affective process, i.e., positive emotion (\texttt{posemo}), negative emotion (\texttt{negemo}) and its subcategories (anxiety (\texttt{anx}), \texttt{anger} and sadness (\texttt{sad})); (2) selected subcategories in the social processes, i.e., \texttt{family} and \texttt{friends}; (3) selected subcategories in the biological processes, i.e., \texttt{body} and \texttt{health}; (4) all subcatgories in the personal concerns, i.e., \texttt{work}, \texttt{leisure}, \texttt{home}, \texttt{money}, religion (\texttt{relig}), and \texttt{death}; and (5) swear words (\texttt{swear}) from informal language, resulting a total of $19$ variables. 
% \textcolor{red}{
We define the model as:
\begin{equation*}
\begin{split}
    \text{Engagement Rate} = \beta_0 &+ \sum_{L_j \in LIWC}(\beta_{j}L_{j}) \\
     &+  \sum_{T_i \in \text{Topics}}(\beta_{i}T_i) + \beta_cF + \epsilon,
\end{split}
\end{equation*}
where $L_j$ and $\beta_j$ are selected LIWC categories and their coefficients.
The topic indicator, $T_i$, and follower count, $F$, are included as control variables. 
% }

Table~\ref{tab:liwc_regression} shows the coefficient regression analysis. Similar results were observed without the controls. 
We focus on the results using {\it Diff} as the dependent variable.
Recall that the {\it non-gcl} audience counts are inherently larger than the {\it gcl} counts. 
The counts are proportional across authors, i.e., an author with a larger {\it non-gcl} audience is more likely to have a larger {\it gcl} audience. %% (although still smaller than their own {\it non-gcl} audiences) compared to other authors.
The results show that using {\it Diff}
cancels out the effect of LIWC categories that are significant for both \textit{gcl} and \textit{non-gcl} engagement rates, for e.g., \texttt{posemo}, \texttt{family}.
It allows us to focus on the LIWC categories that are significant to the \textit{gcl} engagement rate. 
%Therefore, we focus on the results where the dependent variable is Diff.

We observe that the coefficients for \texttt{i} and \texttt{anx} are significant and positive, indicating that a $1\%$ increase of word usage in these two categories would increase the difference between \textit{gcl} and \textit{non-gcl} engagement rates by $0.1\%$ and $0.13\%$ respectively (gcl ER is higher). This suggests that {\bf PHEs sharing personal experiences and feelings (especially anxiety) are associated with relatively higher engagement from {\it gcl} audiences than that of the {\it non-gcl} audiences}. 
We also experimented with a mixed-effect model~\cite{lindstrom1988newton} grouped by authors and found similar results (Table \ref{tab:mixed_effect_full} in Appendix \ref{app:full_liwc}).

\paragraph{Qualitative Analysis}
Table~\ref{tab:qualtitative} presents sample tweets and their \textit{gcl} and \textit{non-gcl} replies. 
The author of the first tweet is located in California while the tweet is about an event in Michigan, resulting in a higher \textit{non-gcl} engagement rate. 
The next two examples demonstrate that sharing personal experiences (with first-person pronouns) is associated with relatively higher \textit{gcl} and \textit{non-gcl} engagement rates.
%, compared to the average of $1.89\%$ and $1.76\%$ respectively (Table~\ref{tab:ER}).
%DONE: \textcolor{red}{mention table where this is coming from}. 
However, we note that in both cases, the \textit{gcl} engagement rate is higher than the \textit{non-gcl} rate.
This is consistent with the regression results in Table~\ref{tab:liwc_regression}, given the occurrence of \texttt{i}  (sharing personal experiences or feelings) in the tweets. 
% It is easier to find such personal Tweets when going through masking-related Tweets, compared to vaccine-ratled Tweets.

%DONE: \textcolor{red}{I am not sure I understand the qualitative analysis of the tweets. In the quant analysis we showed that personal experiences produced higher differences between gcl and non-gcl engagement races. Which tweet is personal versus non personal in these examples? The first tweet does not have "i" but it's clearly personal right? }
% \ph{Add some qualitative examples from the google sheet} 
% https://docs.google.com/spreadsheets/d/16fAnPvLdQvDPReaVeCXbmnMeH0vypXeiH1b7PJwRpB0/edit?usp=sharing

\begin{table*}[!htb]
\scriptsize
\centering
\begin{tabular}{>{\hspace{0pt}}m{0.29\linewidth}>{\hspace{0pt}}m{0.07\linewidth}>{\hspace{0pt}}m{0.09\linewidth}>{\hspace{0pt}}m{0.20\linewidth}>{\hspace{0pt}}m{0.23\linewidth}} \toprule
Tweet - {\it Author state}                                                                                                                                                                                                                          & \textit{gcl} ER  & \textit{non-gcl} ER & \textit{gcl} reply - {\it participant state}                                                                                                      & \textit{non-gcl} reply - {\it participant state}                                                                                   \\ \midrule
2 emerging US surges: Michigan (B.1.1.7) and Northeast (B.1.1.7+ B.1.526). Both variants are vaccine responsive. They need very aggressive vaccination efforts + avoid relaxing mitigation.  -- {\it CA}               &  0.7\%     &  1.8\% & The little inflection that is not receiving adequate attention [URL] -- {\it CA}                                                    &   Michigan has a comprehensive VAX plan in place. [url] -- {\it MI}                                                                                              \\ \midrule
Anybody have ideas for pandemic-era holiday or birthday gifts? \underline{I'm} thinking books, obviously, and magazine subscriptions. Gift certificates to nearby restaurants with delivery or safe take-out? Fun masks -- to grim? Outdoor gear? [url] -- {\it DC}& 42.9\%                                                            & 5.7\%                                                                 & Gift certificates to nearby local movie theatres that are streaming like @[user] -- {\it DC}                                  &      Flowbee, Netflix subscription, Zoom subscription.... Oh wait, do you mean for me or other people? -- {\it WA}                                                                                         \\ \midrule
Had another one of \underline{my} ``\underline{I'm} in a crowded indoor place where no one is masked and then \underline{I} suddenly remember Covid'' dreams last night. How many of you have those too? -- {\it NY}                                                            & 41.7\%                                                            & 38.3\%                                                                & All the time. But I genuinely wonder in the dream, “wait. Is ‘it’ over so it’s ok no one around me is masked??”   -- {\it NY}   & I’ve had about 15 ... no joke! Horror films in my head running for exits through unmasked hoards. -- {\it CA}  \\ \bottomrule
\end{tabular}
\caption{Sampled tweets and replies with \textit{gcl} and \textit{non-gcl} engagement rate (ER). First-personal pronouns in the tweets that indicates self-disclosure are underlined.}
\label{tab:qualtitative}
\end{table*}

\paragraph{LLM-based Content Analysis}
We employed an approach that leveraged LLMs to identify distributional differences between two text corpora \cite{zhong2022describing,zhong2023goal}.
As one option to create two corpora, we used the \textit{Diff} metric to select original tweets in the top $20\%$ (higher \textit{gcl} engagement) and the bottom $20\%$.
For this comparison, we randomly sampled 1000 tweets from each corpus to serve as an exploration (training) set and another 1000 tweets per corpus as the validation set.
we used \texttt{GPT4o-2024-08-06} as a ``proposer'' LLM, which was prompted with sampled tweets to generate $k=20$ hypotheses describing how the two corpora differ.
Each hypothesis was then evaluated by a separate LLM, which estimated a validation score, $V'$. 
We adopted the T5 model with 3B parameters that are fine-tuned for judging whether a hypothesis is true given a text sample.\footnote{https://huggingface.co/ruiqi-zhong/d5\_t5\_validator\_3B}
A positive $V'$ indicates that the hypothesis is more prevalent in the top $20\%$ corpus
(higher \textit{gcl} engagement).
The absolute value $|V'|$ reflects the strength of this difference. 
We report on these results in Table \ref{tab:diff_cf_p20_short}.

To validate the reliability of the $V'$ score, one author annotated a random sample of $100$ tweet-hypothesis pairs. The agreement between the human judgments and the validator LLM was strong, with an accuracy of $96\%$ and a Cohen's Kappa of $0.896$.

From Table \ref{tab:diff_cf_p20_short}, hypotheses with the highest positive values of $V'$ indeed share personal experiences and contain emotional expressions.
Thus, the complementary LLM-based hypotheses are aligned with the results of our prior regression analyses and serve to confirm and validate our results.

To ensure the robustness of the findings,
we experimented with prompt variations that involve different levels of contextual information in the prompts.
Table~\ref{tab:diff_cf_p20_short} uses a minimal-context prompt shown in Table~\ref{tab:prompt_f}, while
\citet{zhong2023goal} adopts a more detailed prompt (Table~\ref{tab:prompt_d5}) that includes explicit information about the corpora. However, we observed that when context about {\it gcl} is included, the LLM tends to hallucinate references to ``locality'' (Table~\ref{tab:diff_p20}).\footnote{We consider referring to locality as hallucination because the co-location between users and the mentioned location in the tweets cannot be inferred from the provided tweets in the prompts.}
When we encourage the model to focus on linguistic and psychological patterns (prompt shown in Table~\ref{tab:prompt_new_target}), the overall content and distributional patterns remained similar.

We also considered additional options to create the two corpora, including varying the metrics used to define the groups, e.g., \textit{gcl}, \textit{non-gcl}, etc., and adjusting the thresholds.
The dominant hypotheses continued to emphasize the sharing of personal experience and emotional expression when the threshold is $10\%$.
Finally, the comparison between \textit{gcl}-based and \textit{non-gcl}-based groupings echoes patterns observed in our regression analysis (Table~\ref{tab:liwc_regression}).
Details and additional results are in Appendix~\ref{app:hypothesis}.

While the LLM-based approach may provide nuanced insights, there are several limitations and concerns that do not arise with our prior statistical approaches and models. 
First, LLMs can hallucinate content.
An example is that the LLM referenced \texttt{local} aspects in a hypothesis, as introduced earlier.
We observe that this issue is largely mitigated when contextual information about \textit{gcl} is excluded from the prompt. 
Although such hallucination is relatively easy to identify, other hypotheses and their associated validation scores may be more difficult to assess for accuracy and reliability.
Further, the granularity and phrasing of hypotheses can be inconsistent.
For instance, while the model often detects emotional tone, it rarely distinguishes between positive and negative sentiment. 
Hypotheses can range from being too vague to being overly specific, and they are often redundant.
These limitations underscore that {\bf while LLM-based methods may provide insights that go beyond the simpler LIWC features, these methods cannot be used in lieu of careful validation and traditional modeling.}

{
\renewcommand{\arraystretch}{1.3} % adds vertical padding between rows
\begin{table}[ht]
\centering
\scriptsize
\begin{tabularx}{\linewidth}{>{
\arraybackslash}X r}
\toprule
\textbf{Hypothesis} & $V'$ \\
\midrule
include expressions of personal emotions or sentiments, such as fear, pride, or frustration & 0.081 \\
focus on personal experiences and anecdotes related to the pandemic & 0.070 \\
discuss everyday life adjustments and challenges during the pandemic, like dating or travel & 0.061 \\
incorporate personal milestones or special occasions and how they've been affected by the pandemic & 0.055 \\
... \\
mention trends or statistics related to COVID-19 cases or immunity & -0.061 \\
include updates or opinions about scientific research and data & -0.120 \\
\bottomrule
\end{tabularx}
\caption{LLM-generated hypotheses with $|V'| \geq 0.05$.  A positive $V'$ score indicates that the hypothesis is more prevalent in the top $20\%$ corpus
(higher \textit{gcl} engagement), compared to the bottom 20\% corpus. Full results in Appendix \ref{app:hypothesis}.}
\label{tab:diff_cf_p20_short}
\end{table}
}

\begin{table}[!htb]
\begin{center}
\scriptsize
\begin{tabular}{lrrr}
\toprule
                         & \textit{Diff}    & \textit{gcl} & \textit{non-gcl}  \\
\midrule
\multicolumn{4}{c}{Selected LIWC categories}                   \\ \midrule
{\bf \texttt{i}}                        & 0.102*** & 0.420***    & 0.318***        \\
\texttt{posemo}                   & 0.038    & 0.087***    & 0.049***        \\
\texttt{negemo}                   & -0.070   & 0.012       & 0.082**         \\
{\bf \texttt{anx}}                      & 0.128*   & -0.052      & -0.179***       \\
\texttt{family}                   & 0.104    & 0.236***    & 0.132**         \\
\texttt{death}                    & 0.020    & -0.041      & -0.061**        \\ \midrule
\multicolumn{4}{c}{Control variables}                   \\ \midrule
\# followers               & -0.001   & 0.011       & 0.012*          \\
masking\_True              & 0.004    & 0.004       & 0.000           \\
education\_True            & -0.001   & -0.001      & -0.000          \\
lockdowns\_True            & -0.003   & -0.003      & -0.000          \\
{\bf vaccines\_True}       & -0.004** & -0.005***   & -0.001          \\
R-squared                  & 0.002    & 0.024       & 0.031           \\
R-squared Adj.             & 0.000    & 0.023       & 0.029           \\
\bottomrule
\end{tabular}
\end{center}
\caption{LIWC regression results of LIWC categories' impact on different types of engagement rates. We only show LIWC categories that are significant in at least one of the regression models (columns). * $p<.1$, ** $p<.05$, ***$p<.01$. Full table in Appendix~\ref{app:full_liwc}.}
\label{tab:liwc_regression}
\end{table}

\subsection{How do \textit{gcl} replies differ from \textit{non-gcl} ones?}
\label{sec:reply}
Our linguistic analysis has focused on the original Tweets. Now, we shift to the replies and attempt to understand the differences in language usage between \textit{gcl} replies and \textit{non-gcl} replies.
% \textcolor{red}{
We first do a sentiment analysis and find \textit{gcl} replies are slightly more positive and less negative than \textit{non-gcl} replies (details in Appendix \ref{app:full_liwc}).
We then run a series of regression analyses to discover additional insights into the differences between {\it gcl} and {\it non-gcl} replies by analyzing the effect of LIWC features on \textit{gcl} and \textit{non-gcl} engagement. 
% }

\begin{figure*}
    \centering
    \includegraphics[scale=0.3]{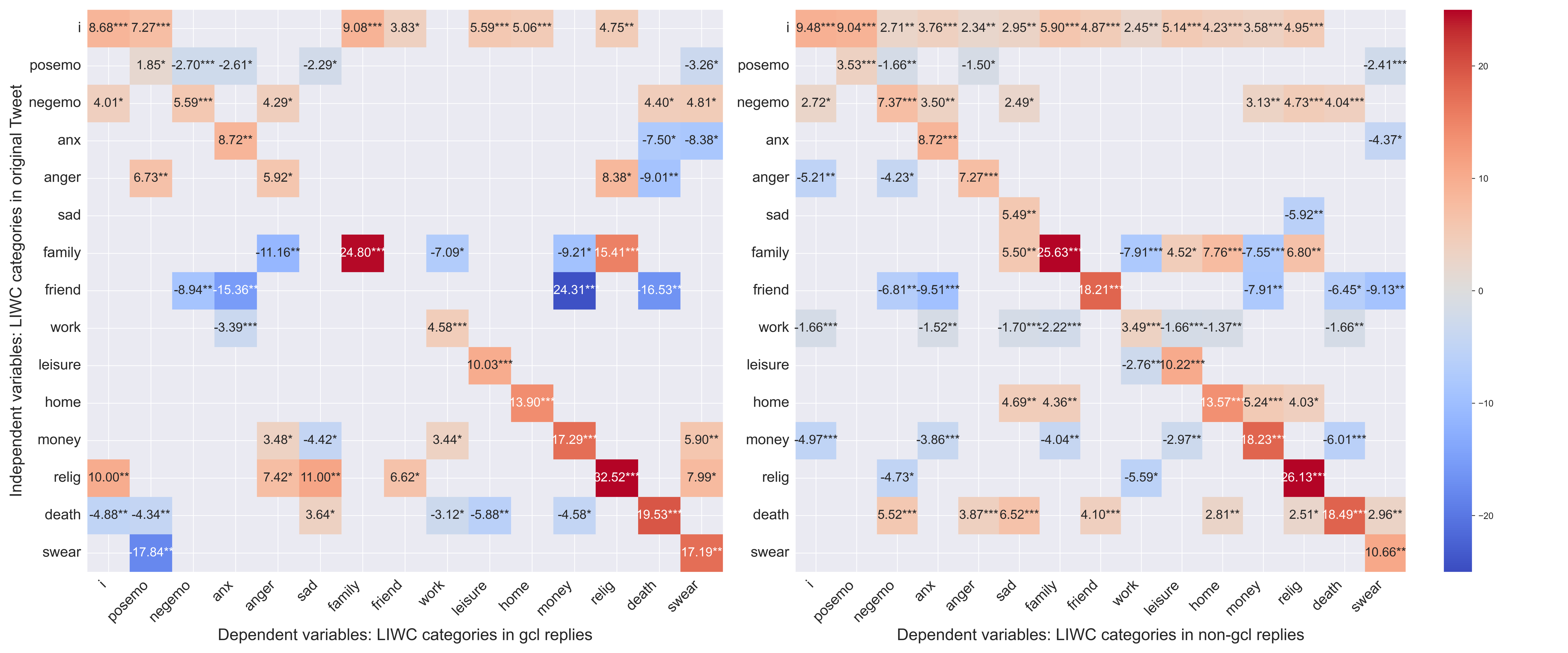}
    \caption{How \textit{gcl} replies and \textit{non-gcl} with specific LIWC categories correlate with LIWC features in the original Tweet. Each column is a logistic regression. Each cell is the coefficient. * $p<.1$, ** $p<.05$, ***$p<.01$. Only selected LIWC categories are shown for demonstration purposes.}
    \label{fig:heatmap}
\end{figure*}

%DONE: \textcolor{red}{what about neutral and negative sentiments? no analysis?}
\paragraph{From Original Tweets to Replies} 
We extend our analysis by investigating how specific language (LIWC categories) in the original tweets might elicit similar language (LIWC categories) in their corresponding \textit{gcl} and \textit{non-gcl} replies. 
To carry out this analysis, we run logistic regressions with the 
independent variables being all the LIWC categories in the original Tweet $T$, and the
dependent variable being one binarized LIWC category present in the replies for that tweet $T$.\footnote{A reply-level model would violate the assumption of independent observations, as all replies to a single tweet share identical independent variables.} Formally, the model is defined as:

\begin{equation*}
L_{i}^{\text{gcl\_replies}} = \beta_0 + \sum_{L_{i}^T \in LIWC}(\beta_i L_{i}^T) + \epsilon,
\end{equation*}
where $L_{i}^{\text{gcl\_replies}}$ is a binary value of $1$ if the $i$-th LIWC category is present in at least one of the \textit{gcl} replies to $T$, and $L_{i}^T$ is the $i$-th LIWC score of $T$. We run this model for each binarized LIWC category (dependent variable), and for 
\textit{gcl} and \textit{non-gcl} replies, separately.
A comparison of the regression coefficients for the \textit{gcl} and \textit{non-gcl} groups (see Figure~\ref{fig:heatmap}) reveals a strong diagonal effect for both groups, with slightly larger diagonal coefficients in \textit{gcl} replies for most personal concerns (\texttt{work}, \texttt{home}, \texttt{relig}, and \texttt{death}) and swear words. 
This suggests that \textit{gcl} replies may be more likely to discuss personal concerns compared to \textit{non-gcl} replies, when these concerns have been raised in the original Tweets. However, there is a stronger agreement in \textit{non-gcl} replies for LIWC categories associated with affective processes, including \texttt{posemo} and \texttt{negemo} and all its subcategories, \texttt{anx}, \texttt{anger}, and \texttt{sad}.
This suggests that, while \textit{gcl} and \textit{non-gcl} replies discuss similar types of personal concerns when they are mentioned in the original Tweets, \textit{non-gcl} replies are more likely to share similar emotions to the original Tweets. 
%DONE: \textcolor{red}{are you sure the appropriate word is affection?}
We hypothesize that this may be due to \textit{non-gcl} participants having weaker social connections with the authors, leading them to share similar emotions and potentially avoiding conflict, while \textit{gcl} participants express more freely a diverse set of emotions independently of the ones that are raised in the original Tweets from local authors. 

More interestingly, when observing the heatmaps horizontally, the coefficients for \texttt{i} are significant and positive in almost all regressions for \textit{non-gcl} replies, while only significant and positive in \texttt{posemo}, social processes (\texttt{family} and \texttt{friend}), and some personal concerns (\texttt{leisure}, \texttt{home}, \texttt{relig}) for \textit{gcl} replies. 
This pattern indicates that \textbf{authors sharing personal experiences tend to elicit all kinds of psychological and emotional states from \textit{non-gcl} replies but mostly positive \textit{gcl} replies. This finding reveals that geo-co-located messaging could be used as a strategy for enhancing engagement and maintaining a positive discourse in online communities.}

\paragraph{Difference in Replies} 
To examine differences in language usage between \textit{gcl} and \textit{non-gcl} replies, we propose to run a logistic regression where the dependent variable indicates whether a reply is \textit{gcl} or \textit{non-gcl} and the independent variables are the LIWC categories described in Section~\ref{sec:tweet}. A coefficient analysis will allow us to identify linguistic features that are uniquely associated with \textit{gcl} or \textit{non-gcl} replies.
As the Table~\ref{tab:logit} shows, negative emotions (particularly anger) are more prevalent in \textit{non-gcl} replies;
while the use of pronouns such as \texttt{we} and \texttt{you} is more common in \textit{gcl} replies, which might suggest more direct communication with the \textit{gcl} PHE authors. 
Interestingly, \textit{gcl} and \textit{non-gcl} replies are associated with different types of personal concerns. For example, \textit{gcl} replies involve topics related to \texttt{work}, \texttt{home}, and \texttt{money} more frequently, while \textit{non-gcl} replies are more likely to discuss \texttt{relig} and \texttt{death} and biological processes (i.e., \texttt{body} and \texttt{health}).

\begin{table}[tb]
\centering
\scriptsize
\begin{tabular}{@{}lcclcc@{}} \toprule
Variable & OR & 95\% CI & Variable & OR & 95\% CI \\ \midrule
i        & 0.63*     & (0.39, 1.03)    & body     & 0.14*** & (0.05, 0.39)    \\
we       & 4.25*** & (2.15, 8.40)    & health   & 0.27*** & (0.14, 0.52)    \\
you      & 7.73*** & (5.12, 11.68)   & work     & 2.08*** & (1.38, 3.13)    \\
posemo   & 1.04     & (0.78, 1.40)    & leisure  & 0.86     & (0.37, 1.99)    \\
negemo   & 0.39*** & (0.20, 0.76)    & home     & 44.48*** & (12.12, 163.25) \\
anx      & 0.60     & (0.16, 2.20)    & money    & 3.85*** & (1.78, 8.31)    \\
anger    & 0.24*** & (0.08, 0.69)    & relig    & 0.29* & (0.11, 0.82)    \\
sad      & 2.01     & (0.73, 5.54)    & death    & 0.17*** & (0.06, 0.50)    \\
family   & 0.25*     & (0.06, 1.06)    & swear    & 1.93     & (0.74, 4.98)    \\
friend   & 0.26     & (0.05, 1.30)    &          &          &                 \\ \bottomrule
\end{tabular}
\caption{Logistic regression results for \textit{gcl} vs. \textit{non-gcl} replies. Odds Ratios (OR) over $1$ indicate a category is more frequent in \textit{gcl} replies.* $p<.1$, ** $p<.05$, ***$p<.01$.}
\label{tab:logit}
\end{table}

% \textcolor{red}{
We applied the LLM-based method mentioned in Section \ref{sec:tweet} to examine distributional differences between \textit{gcl} and \textit{non-gcl} replies. 
As shown in Table~\ref{tab:gcl_replies}, however, all generated hypotheses yielded $V'$ scores below $0.05$.  
This indicates that the method did not uncover any strong or consistent distributional distinctions between the two reply types, unlike the LIWC-based analysis, which revealed clear linguistic and topical divergences.
% }

\subsection{Which PHEs have more \textit{gcl} engagement?}
\label{sec:user}
This analysis shifts the focus to PHE authors and explores PHE features that may attract more \textit{gcl} engagement.

\begin{table}[t]
\centering
\small
\begin{tabular}{lrrrr} \toprule
Professions & Media   & Academia & Health & Political  \\ \midrule
\textit{gcl} ER      & 1.63\%  & 2.09\%   & 1.95\%  & 1.72\%     \\
\textit{non-gcl} ER  & 1.55\%  & 1.89\%   & 1.73\%  & 1.65\%     \\
{\bf \textit{Diff}}        & {\bf 0.08\%} & {\bf 0.20\%}  & {\bf 0.23\%} & {\bf 0.07\%}    \\ 
{\bf IRR}  & {\bf 0.923} & {\bf 1.325} & 1.013   & {\bf 2.153}   \\
\# authors  & 131     & 99       & 108     & 20         \\
\# Tweets   & 9107    & 5610     & 6734    & 1394       \\
Test Acc.   & 88\%    & 98\%     & 94\%    & 98\%            \\ \bottomrule
\end{tabular}
\caption{Differences between \textit{gcl} and \textit{non-gcl} engagement rates (ER) for tweets grouped by author profession. IRR are results from Poisson mixed-effects regression (Eq. \ref{eq:poisson_mixed}); a value greater than $1$ indicates a higher ER for the {\it gcl} group. Test Acc is \texttt{gpt4}'s accuracy on a sample of 50 authors. Statistically significant results ($p < 0.001$) are shown in bold.}
\label{tab:er_professions}
\end{table}

\paragraph{Profession} The PHE authors provide detailed user profiles and generate substantial content.
Similar to the use of LLMs to understand social context \cite{ziems2024can}, we use \texttt{gpt4} to predict the authors' professions. 
Based on a preliminary manual review of a sample of author profiles, we selected four broad professional categories relevant to our analysis:
media, academia, health, and political/policy.
For each group, we included example professions in the prompt, e.g., journalists and reporters for the media profession. 
LLM generates a binary prediction for each profession category and the rationale for the prediction.
The complete prompt, sample data and predictions are in Appendix \ref{app:prompt}.
The authors manually reviewed the classification results for $50$ authors with above $90\%$ accuracy for all four professions.

The tweet-level analysis in Table \ref{tab:er_professions} shows consistently higher \textit{gcl} engagement rates across all professions. 
However, applying the mixed-effects Poisson regression (Eq. \ref{eq:poisson_mixed}) separately to the tweets from each professional group, we observe that the relationship between geo-co-location and engagement varies by PHE's profession.
The strength of {\it gcl} engagement is the highest for political professionals (IRR $= 2.153$), followed by academic professionals ($1.325$).  Media professionals are associated with significant but slightly lower {\it gcl} engagement ($0.923$), while health professionals show no significant difference ($1.013$).

\section{Summary, Lessons Learned, Limitations and Future Work}
\label{sec:limitation}

\paragraph{Summary}

This study examines the role of geographic co-location in shaping social media engagement during the COVID-19 pandemic.
Using Poisson regression models and statistical tests, we find that users who are \textit{gcl} with PHEs are associated with higher engagement rates
To explore the nature of this engagement, we analyzed conversational content, including topics, language use, and the characteristics of PHEs. We observe that emotional language and the sharing of personal experiences are associated with increased engagement from \textit{gcl} participants, who more frequently express positive sentiment and personal issues in their replies. 
% The relationship between \textit{gcl} and engagement also appears stronger for PHEs in academic and medical professions compared to those in media or political roles.
However, this relationship varies by PHE's professions. Our mixed-effects analysis reveals a positive association with {\it gcl} engagement for academic and political professionals, a modest negative association for media professionals, and non-significance for those in medical roles.
We additionally applied an LLM-based method to generate hypotheses about content differences between tweet groups. Many of the resulting hypotheses, such as the emphasis on personal experience and emotional expression, were consistent with patterns observed in our regression analysis. However, limitations in specificity and reliability suggest that such methods should be interpreted with caution and warrant further refinement.
Overall, these findings indicate that geographic co-location is an important factor to consider in the study of public health communication. Content that is locally resonant and personally expressive may be associated with higher engagement, offering possible directions for future research on effective messaging strategies in online public health discourse.

\paragraph{Lesson learned for public health messaging}
A key insight is that PHEs' effort to share personal experiences and use positive emotional language tends to receive higher \textit{gcl} engagement and replies with more positive sentiment.
This is an encouraging insight given that the overall COVID-19 discourse on social media is largely negative.
While we have not attempted to explore causality, these findings suggest potential for PHEs to improve engagement by leading more localized discussions, linking guidance to community resources, and personalizing messages. Although PHEs cannot control their audience composition, they can take steps to strengthen connections with their \textit{gcl} audience and foster more effective public health communication.

While the focus of this dataset was interactions between PHEs and the public during COVID-19, 
the analysis and models may be applied to other scenarios. 
Related work has studied engagement dynamics and polarization in the context of abortion and political discourse \cite{lerman2024affective,Nettasinghe2025}. 
Research in \cite{rao2021political} explored a national index of beliefs and attitudes around evidence based findings and science. Ongoing work aims to develop a national index of attitudes and emotions about community challenges including homelessness, substance abuse disorder, emotional health, etc.
The hope is that such national indices, combined with findings on improved engagement, can lead to better approaches for PHEs and community decision makers to engage with the public.

\paragraph{Limitations and future work} 
Twitter users who use location services may be more active \cite{rzeszewski2017spatial}, which may introduce a bias. 
Findings derived from Twitter data may not generalize to other platforms or to populations that are less active or less visible on social media.
This sampling bias may result in unequal targeting of messages, with engagement-optimized strategies disproportionately focusing on more vocal or digitally connected groups.

Additionally, users may provide inaccurate location information~\cite{hecht2011tweets} and the geo-location inference tool may be inaccurate. 
Our analysis uses state-level geo-location, a choice made to ensure data reliability, as the accuracy of currently available geo-inference tools degrades substantially at finer granularities~\cite{zhang2022changes,masis2024earth}.
The accuracy can be improved by using advanced geo-inference tools leveraging deep learning models~\cite {zhang2023geo}.

% or by including richer user data - content and the social network~\cite{tian2020twitter}.
LIWC and the employed LLM-based method provided complementary insights, though both have limitations, i.e., LIWC in the scope of categories it captures, and the LLM method in consistency and specificity.
As this is a rapidly evolving field~\cite{zhong2024explaining,movva2025sparse}, these LLM-based methods should be interpreted with caution. 
While the natural language hypotheses are potentially more expressive than fixed LIWC categories, they can lack the precision of human-expert-curated ones, sometimes resulting in vague or imprecise statements.
Furthermore, although the validator model showed strong agreement with human annotation on our test sample, its performance may not generalize to more complex or ambiguous text-hypothesis pairs. 
Finally, the method is purely descriptive. It is adept at identifying correlational patterns but cannot provide causal insight into why these differences in language emerge. Future work is needed to improve the specificity, robustness, and causal inference capabilities of these novel techniques.

There is also a risk that insights about effective message framing, such as the use of personal narratives or emotional language, could be misused to manipulate public sentiment rather than inform it.
Future work should evaluate the impact of geo-co-location in other online contexts (e.g., Reddit), and expand content analysis to other domains beyond public health to assess the broader applicability and ethical considerations of this approach.
% The impact of geo-co-location should also be explored using datasets from other platforms such as Reddit, and on the discourse beyond the public health domain.

% % \begin{comment}
% Users who are \textit{gcl} with PHEs exhibit higher engagement rates in discussions about masking, lockdowns, and education. Conversely, vaccine discussions attracted higher \textit{non-gcl} engagement.
% Emotional language and personal experiences significantly enhance engagement from \textit{gcl} participants; the replies express  positive sentiment. 
% The \textit{gcl} participants express more positive emotions and personal issues in response to tweets that share personal experiences or emotions.
% There is greater \textit{gcl} engagement for PHEs in academic and medical fields compared to media and political professions.
% \end{comment}

\section{Acknowledgments}

This work was supported by the National Science Foundation under Grant No. CCF-2200256.

\bibliography{reference}

\section*{Checklist}

\begin{enumerate}

\item For most authors...
\begin{enumerate}
    \item  Would answering this research question advance science without violating social contracts, such as violating privacy norms, perpetuating unfair profiling, exacerbating the socio-economic divide, or implying disrespect to societies or cultures?
    \answerYes{Yes}
  \item Do your main claims in the abstract and introduction accurately reflect the paper's contributions and scope?
    \answerYes{Yes}
   \item Do you clarify how the proposed methodological approach is appropriate for the claims made? 
    \answerYes{Yes}
   \item Do you clarify what are possible artifacts in the data used, given population-specific distributions?
    \answerYes{Yes, in Section~\ref{sec:limitation}}.
  \item Did you describe the limitations of your work?
    \answerYes{Yes, in Section~\ref{sec:limitation}}.
  \item Did you discuss any potential negative societal impacts of your work?
    \answerYes{Yes, in Section~\ref{sec:limitation}.}
      \item Did you discuss any potential misuse of your work?
    \answerYes{Yes, in Section~\ref{sec:limitation}}.
    \item Did you describe steps taken to prevent or mitigate potential negative outcomes of the research, such as data and model documentation, data anonymization, responsible release, access control, and the reproducibility of findings?
    \answerYes{Yes}
  \item Have you read the ethics review guidelines and ensured that your paper conforms to them?
    \answerYes{Yes}
\end{enumerate}

\item Additionally, if your study involves hypotheses testing...
\begin{enumerate}
  \item Did you clearly state the assumptions underlying all theoretical results?
    \answerYes{Yes, we have some repeating hypotheses testing with a similar setup. We clearly state the assumption for the first test and briefer statements for the rest for readability.}
  \item Have you provided justifications for all theoretical results?
    \answerYes{Yes}
  \item Did you discuss competing hypotheses or theories that might challenge or complement your theoretical results?
    \answerYes{Yes}
  \item Have you considered alternative mechanisms or explanations that might account for the same outcomes observed in your study?
    \answerYes{Yes}
  \item Did you address potential biases or limitations in your theoretical framework?
    \answerYes{Yes, in Section~\ref{sec:limitation}}
  \item Have you related your theoretical results to the existing literature in social science?
    \answerYes{Yes}
  \item Did you discuss the implications of your theoretical results for policy, practice, or further research in the social science domain?
    \answerYes{Yes, in Section~\ref{sec:limitation}}
\end{enumerate}

\item Additionally, if you are including theoretical proofs...
\begin{enumerate}
  \item Did you state the full set of assumptions of all theoretical results?
    \answerNA{NA}
	\item Did you include complete proofs of all theoretical results?
    \answerNA{NA}
\end{enumerate}

\item Additionally, if you ran machine learning experiments...
\begin{enumerate}
  \item Did you include the code, data, and instructions needed to reproduce the main experimental results (either in the supplemental material or as a URL)?
    \answerNA{NA}
  \item Did you specify all the training details (e.g., data splits, hyperparameters, how they were chosen)?
    \answerNA{NA}
     \item Did you report error bars (e.g., with respect to the random seed after running experiments multiple times)?
    \answerNA{NA}
	\item Did you include the total amount of compute and the type of resources used (e.g., type of GPUs, internal cluster, or cloud provider)?
    \answerNA{NA}
     \item Do you justify how the proposed evaluation is sufficient and appropriate to the claims made? 
    \answerNA{NA}
     \item Do you discuss what is ``the cost`` of misclassification and fault (in)tolerance?
    \answerNA{NA}
  
\end{enumerate}

\item Additionally, if you are using existing assets (e.g., code, data, models) or curating/releasing new assets, \textbf{without compromising anonymity}...
\begin{enumerate}
  \item If your work uses existing assets, did you cite the creators?
    \answerYes{Yes}
  \item Did you mention the license of the assets?
    \answerNo{NA}
  \item Did you include any new assets in the supplemental material or as a URL?
    \answerNo{No}
  \item Did you discuss whether and how consent was obtained from people whose data you're using/curating?
    \answerNo{No}
  \item Did you discuss whether the data you are using/curating contains personally identifiable information or offensive content?
    \answerNo{No}
\item If you are curating or releasing new datasets, did you discuss how you intend to make your datasets FAIR?
\answerNA{NA}
\item If you are curating or releasing new datasets, did you create a Datasheet for the Dataset? 
\answerNA{NA}
\end{enumerate}

\item Additionally, if you used crowdsourcing or conducted research with human subjects, \textbf{without compromising anonymity}...
\begin{enumerate}
  \item Did you include the full text of instructions given to participants and screenshots?
    \answerNA{NA}
  \item Did you describe any potential participant risks, with mentions of Institutional Review Board (IRB) approvals?
    \answerNA{NA}
  \item Did you include the estimated hourly wage paid to participants and the total amount spent on participant compensation?
    \answerNA{NA}
   \item Did you discuss how data is stored, shared, and deidentified?
   \answerNA{NA}
\end{enumerate}

\end{enumerate}
\appendix
\clearpage
\section{Seed PHEs}
\label{app:phe}
The usernames for the 30 seed PHEs on Twitter are:
EricTopol, PeterHotez, ashishkjha, trvrb, EpiEllie, JuliaRaifman, devisridhar, meganranney,
luckytran, asosin, DrLeanaWen, dremilyportermd, DrJaimeFriedman, davidwdowdy, BhramarBioStat, geochurch, DrEricDing, michaelmina\_lab, Bob\_Wachter, JenniferNuzzo, mtosterholm,
MonicaGandhi9, cmyeaton, nataliexdean, angie\_rasmussen, ProfEmilyOster, mlipsitch, drlucymcbride, ScottGottliebMD, CDCDirector, and Surgeon\_Genera

\section{Supplement Results for Section \ref{sec:engagement}}
\label{app:poisson}
As a robustness check for the Poisson regression presented in Section~\ref{sec:engagement}, we fit two advanced models to account for author-level heterogeneity. 
Both use the same structure as the original model, with each original tweet corresponding to two data points, one for the \textit{gcl} audience and another for \textit{non-gcl}.
Both models confirm that the positive association between \textit{gcl} and engagement is statistically significant and robust. 
The results are summarized in Table \ref{tab:robust_poisson}.

First, we fit a {\bf mixed-effects} Poisson regression. This model includes a random intercept for each PHE author, treating author-level effects as random variables. We define the model as:
\begin{equation*}
    \log(\mathbb{E}[Y_{ij}]) = \beta_0 + \beta_1X_{ij} + u_j + \log(W_{ij})
\end{equation*}
where $Y_{ij}$ is the reply count for tweet $i$ by author $j$ for a specific audience type, $X_{ij}$ is a binary indicator for that audience type (\textit{gcl} vs. \textit{non-gcl}), $u_j$ is the random intercept for author $j$, and $W_{ij}$ is the corresponding audience size included as an exposure variable.

As a supplementary check, we also fit a Poisson regression with author {\bf fixed effects}. This model accounts for author heterogeneity by including a separate fixed intercept for each author. The model is specified as:
\begin{equation*}
    \log(\mathbb{E}[Y_{ij}]) = \alpha_j + \beta_1X_{ij} + \log(W_{ij})
\end{equation*}
where $\alpha_j$ is the unique fixed-effect intercept for author $j$. This model requires sufficient observations per author to reliably estimate each fixed effect. To ensure model convergence, this analysis was restricted to authors with at least three observations.

\begin{table}[h!]
\centering
\resizebox{\linewidth}{!}{
\begin{tabular}{lrrr}
\toprule
Model & Coefficient & IRR & 95\% CI (IRR) \\
\midrule
Mixed-Effects & -0.122 & 0.885*** & (0.870, 0.900) \\
Fixed-Effects & -0.105 & 0.901*** & (0.883, 0.919) \\
\bottomrule
\end{tabular}
}
\caption{Robustness Checks for Poisson Regression. $X_T$ (\textit{gcl} $= 0$, \textit{non-gcl} $= 1$). IRR is the incidence rate ratio. $p < 0.001$ is marked with ***.}
\label{tab:robust_poisson}
\end{table}

\section{COVID-19 Issues}
\label{app:issue}

The lockdown issue comprises content pertaining to early state and federal mitigation efforts, such as quarantines, stay-at-home orders, business closures, reopening, and calls for social distancing. The masking issue is defined by discussions on the use of face coverings, mask mandates, mask shortages, and anti-mask sentiment. Education-related content involves tweets about school closures, reopening of educational institutions, homeschooling, and online learning during the pandemic. The vaccine issue pertains to discussions about COVID-19 vaccines, vaccine mandates, anti-vaccine sentiment, and vaccine hesitancy in the U.S.
\begin{table}[!htb]
\centering
\footnotesize
\begin{tabular}{>{\hspace{0pt}}m{0.195\linewidth}>{\hspace{0pt}}m{0.7\linewidth}} 
\toprule
Issue     & Sample Tweets                                                                                                                                                               \\ \midrule
Lockdowns & This is a GREAT idea. We’re all in this together.
Take care of each other. \textbf{\#StayHome} \#TakeItSeriously \#FlattenTheCurve \#COVID19~ ~                                      \\ \midrule
Masking   & We’re in the middle of a pandemic and y’all are
still coughing and sneezing without \textbf{covering your
mouths}? Come on now.~ ~                                                   \\ \midrule
Education & More glimmers of hope as we “safely” move
forward and open up Texas AM \textbf{University} while
containing \#COVID19.                                                               \\ \midrule
Vaccines  & You are joking right? Zero sympathy for \textbf{antivaxxers} who quit their jobs rather than get \textbf{vaccinated}. They put us all at risk and make the
pandemic prolonged for the world.  \\
\bottomrule
\end{tabular}
\caption{Sample Tweets for each COVID-19 issue. Issue-relevant keywords are bolded. Table from \citet{rao2023pandemic}.}
\label{tab:sample_issue}
\end{table}

\section{Lexical Analysis for Geo-co-located Tweets}
\label{app:lexical}

% To validate that the proposed engagement rate captures meaningful Tweets, 
To gain insights into what lexical features are most distinctive for \textit{gcl} tweets,
we compare Tweets that have 
$0\%$ \textit{gcl} engagement rate ($N=10,854$, bottom $~68\%$ of the tweets) with the top $10\%$ percentile of tweets 
($3.4\%$ \textit{gcl} engagement rate with $N=1,576$).We also compare 
$0\%$ \textit{gcl} ER tweets with the top $20\%$ percentile of tweets ($1.0\%$ \textit{gcl} engagement rate with $N=3,190$).

\begin{figure}[!htb]
    \centering
    \begin{subfigure}[b]{0.45\textwidth}
        \centering
        \includegraphics[width=\columnwidth]{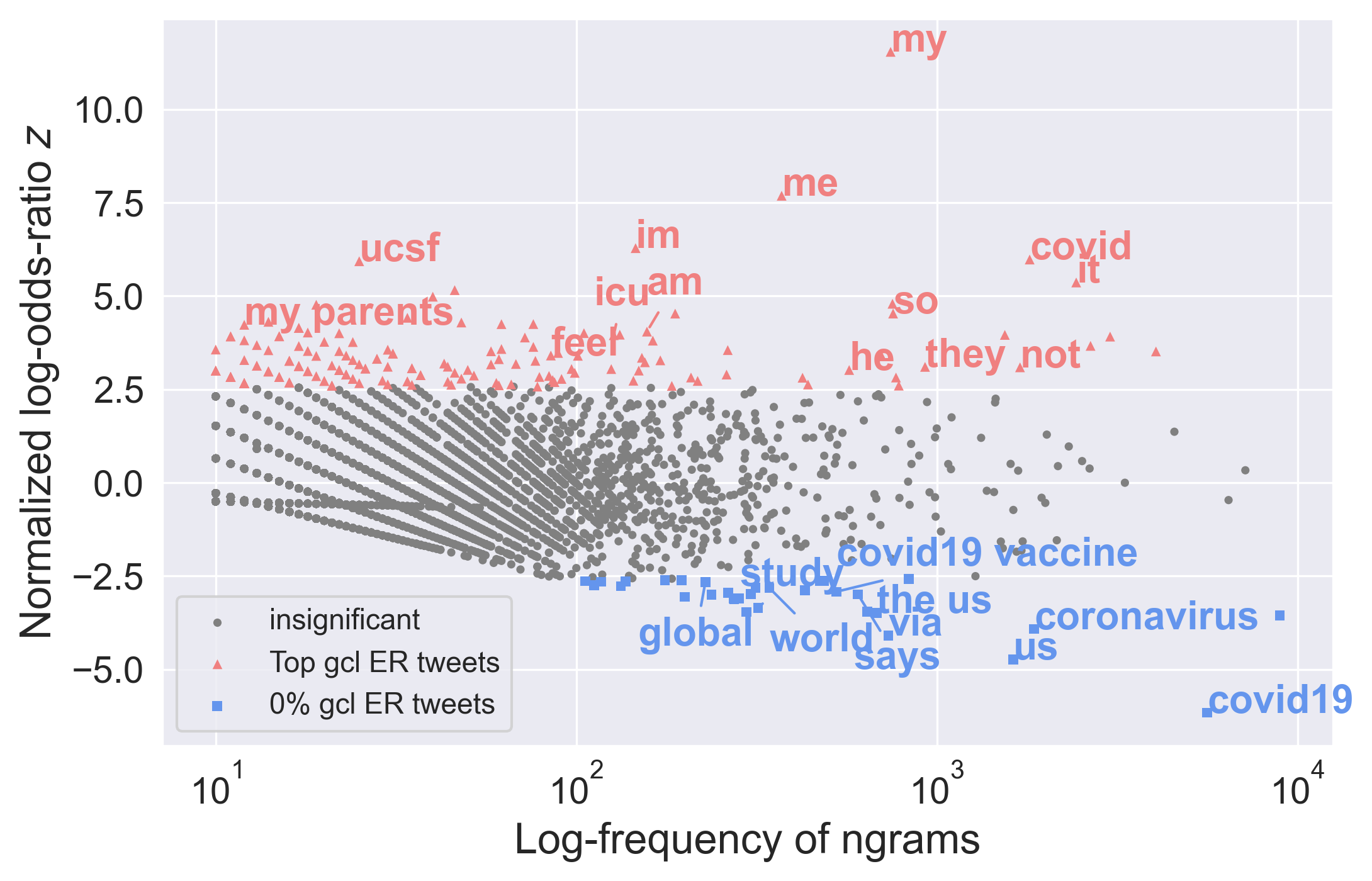}
        \caption{Top $10\%$ percentile of tweets}
    \end{subfigure}
    \hfill
    \begin{subfigure}[b]{0.45\textwidth}
        \centering
        \includegraphics[width=\columnwidth]{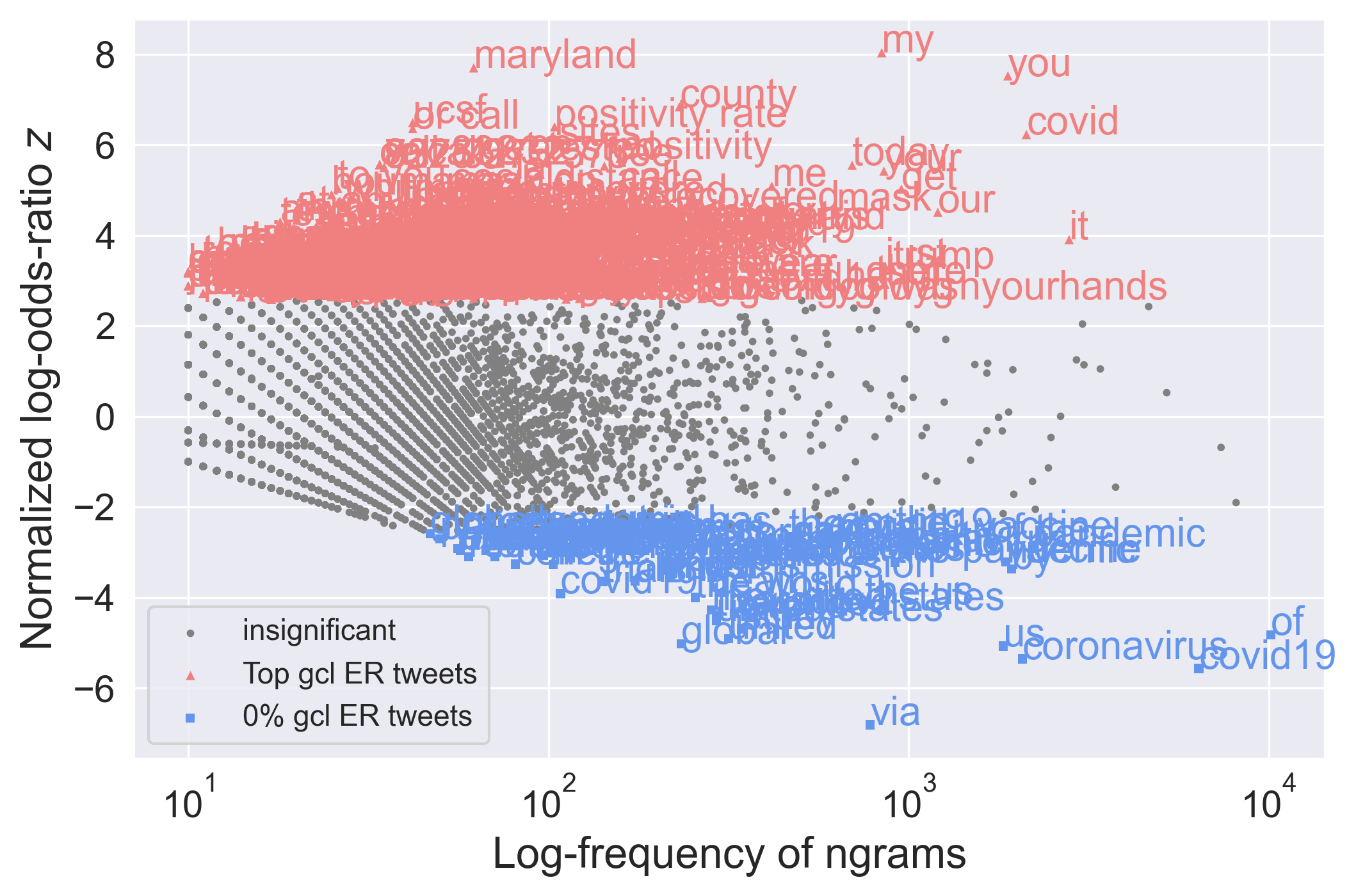}
        \caption{Top $20\%$ percentile of tweets}
    \end{subfigure}
    \caption{Most distinctive n-grams ($n\in \{1,2,3\}$): tweets with $0\%$ \textit{gcl} engagement rate (bottom $~68\%$ of the tweets) vs. (a) the top $10\%$ percentile of tweets ($3.4\%$ \textit{gcl} engagement rate) and (b) the top $20\%$ percentile of tweets ($1.0\%$ \textit{gcl} engagement rate) . Only n-grams with $99.5\%$ confidence level are colored ($z > 2.57$)}
    \label{fig:lexical}
\end{figure}

We follow \citet{monroe2008fightin} and use normalized log-odds-ratio $z$ to find out the n-grams ($n\in \{1,2,3\}$) more associated with Tweets with higher and lower \textit{gcl} engagement rates. 
We find that the top group shares more personal experiences/feelings (``my'', ``me''), 
% They have keywords like “my parents” get buried in the plot.
while the bottom group talks about the national or global situation of COVID-19 with keywords such as “covid19”, “us”, “united states”, “study”, “the world” (Figure~\ref{fig:lexical}).

% top 20%: ER = 1.0%, n = 3190
% top 30%: ER = 0.2%, n = 4746

% Furthermore, we differentiate the proposed measure with a simpler measure that only considers the percentage of GCL replies of a Tweet. 
% We compare Tweets whose replies have over $33\%$ GCL replies ($N=1,641$, top $10\%$ percentile) versus Tweets whose replies have $0\%$ GCL replies ($N=10,699$, bottom $67\%$ percentiles). 
% We find that the top group contains more specific location information (“county”, “texas”, “la”, “gavinnewsom”etc.) and may combine with hashtags like “washyourhands”, “socialdistancing”, “stayathome”, “justamask”.
% The bottom group shares more national news or comments on the national level (“us”, “trump”, “the us”, “study”, “world”, “global”, “according to”).
% The fact that the top group selects Tweets that have different linguistic features when compared to the ones selected by the proposed measure suggests these two metrics are different. \ph{More on why we want the proposed one? We cannot compare distributions with the simple metrics.}

\section{Supplemental Results for Section \ref{sec:tweet}}
\label{app:full_liwc}

\paragraph{Sentiment} To investigate sentiment expressed in the \textit{gcl} and \textit{non-gcl} replies, we use a RoBERTa-base model \cite{loureiro-etal-2022-timelms} trained on ~124M tweets from January 2018 to December 2021, and fine-tuned for sentiment analysis with the TweetEval benchmark \cite{rosenthal2017semeval}.
We classify each reply as positive, neutral, or negative.
For each original Tweet $T$, we compute the positive rate for its \textit{gcl} replies and \textit{non-gcl} replies, respectively.
We find that \textit{gcl} replies are slightly more positive and less negative than \textit{non-gcl} replies.
The mean percentage of \textit{gcl} replies labeled as positive is $14\%$ and the standard deviation is $29\%$, while the mean percentage for \textit{non-gcl} replies labeled as positive is $13\% \pm 22\%$. The difference between the two distributions was statistically significant (KS test with $p<0.001$ and $n=3,521$). 
The mean percentage of \textit{gcl} and \textit{non-gcl} replies labeled as negative were also significantly different ($p<0.001$) with mean and standard deviation values of $43\% \pm 39\%$ and $45\% \pm 29\%$, respectively.

\paragraph{Regression Results for the Impact of COVID-19 Topics}
We define the linear regression model for directly evaluating the impact of COVID-19 topics on engagement rates as follows:
\begin{equation*}
    \text{Engagement Rate} = \beta_0 + \sum_{T_i \in \text{Topics}}(\beta_{i}T_i) + \beta_cF + \epsilon,
\end{equation*}
where the dependent variable is one of three engagement rates (\textit{gcl}, \textit{non-gcl}, or \textit{Diff}). The independent variables are binary indicators for each topic $T_i$ (masking, lockdowns, education, and vaccines), and $F$ is a control for the author's min-max normalized follower count to account for their general influence.
The regression results are shown in Table \ref{tab:topic_regression}.

\begin{table}
\begin{center}
\scriptsize
\begin{tabular}{lrrr}
\toprule
                      & \textit{Diff}    & \textit{gcl} & \textit{non-gcl}  \\
\midrule
const                  & 0.002**  & 0.020***    & 0.018***        \\
                       & (0.001)  & (0.001)     & (0.001)         \\
masking\_True          & 0.003    & 0.004       & 0.001           \\
                       & (0.003)  & (0.003)     & (0.002)         \\
education\_True        & -0.001   & -0.000      & 0.000           \\
                       & (0.003)  & (0.003)     & (0.002)         \\
lockdowns\_True        & -0.003   & -0.003      & 0.000           \\
                       & (0.005)  & (0.005)     & (0.003)         \\
{\bf vaccines\_True}         & -0.004** & -0.004**    & -0.001          \\
                       & (0.002)  & (0.002)     & (0.001)         \\
\# followers           & -0.004   & -0.031***   & -0.027***       \\
                       & (0.008)  & (0.007)     & (0.006)         \\
R-squared              & 0.000    & 0.002       & 0.002           \\
R-squared Adj.         & 0.000    & 0.001       & 0.001           \\
\bottomrule
\end{tabular}
\end{center}
\caption{Regression results for evaluating the impact of COVID-19 Topics. Standard errors in parentheses. * $p<.1$, ** $p<.05$, ***$p<.01$}
\label{tab:topic_regression}
\end{table}

\paragraph{Complete Results for the LIWC regression}
The complete results of the LIWC regression analysis (a complete version of Table \ref{tab:liwc_regression}) in Section~\ref{sec:tweet} is shown in Table \ref{tab:liwc_regression_full}.

\begin{table}
\begin{center}
\scriptsize
\begin{tabular}{lrrr}
\toprule
                         & \textit{Diff}    & \textit{gcl} & \textit{non-gcl}  \\
\midrule
const                  & 0.001    & 0.017***    & 0.016***        \\
                       & (0.001)  & (0.001)     & (0.001)         \\
{\bf i}                      & 0.103*** & 0.448***    & 0.345***        \\
                       & (0.038)  & (0.037)     & (0.028)         \\
we                     & -0.014   & -0.036      & -0.022          \\
                       & (0.030)  & (0.029)     & (0.022)         \\
you                    & -0.022   & 0.017       & 0.039           \\
                       & (0.039)  & (0.038)     & (0.029)         \\
posemo                 & 0.038    & 0.092***    & 0.054***        \\
                       & (0.024)  & (0.023)     & (0.018)         \\
negemo                 & -0.070   & 0.008       & 0.079**         \\
                       & (0.046)  & (0.045)     & (0.034)         \\
{\bf anx}                    & 0.128*   & -0.053      & -0.181***       \\
                       & (0.069)  & (0.067)     & (0.051)         \\
anger                  & 0.106    & 0.037       & -0.069          \\
                       & (0.073)  & (0.071)     & (0.054)         \\
sad                    & 0.057    & -0.018      & -0.075          \\
                       & (0.074)  & (0.072)     & (0.054)         \\
family                 & 0.105    & 0.248***    & 0.143**         \\
                       & (0.086)  & (0.084)     & (0.063)         \\
friend                 & -0.022   & 0.039       & 0.061           \\
                       & (0.093)  & (0.090)     & (0.069)         \\
body                   & -0.066   & -0.066      & 0.000           \\
                       & (0.055)  & (0.053)     & (0.040)         \\
health                 & -0.017   & -0.014      & 0.002           \\
                       & (0.026)  & (0.025)     & (0.019)         \\
work                   & 0.007    & -0.011      & -0.018          \\
                       & (0.018)  & (0.018)     & (0.014)         \\
leisure                & -0.059   & -0.028      & 0.031           \\
                       & (0.039)  & (0.038)     & (0.029)         \\
home                   & 0.019    & 0.003       & -0.016          \\
                       & (0.060)  & (0.058)     & (0.044)         \\
money                  & -0.035   & -0.050      & -0.015          \\
                       & (0.042)  & (0.040)     & (0.031)         \\
relig                  & -0.008   & -0.079      & -0.071          \\
                       & (0.081)  & (0.079)     & (0.060)         \\
death                  & 0.018    & -0.055      & -0.073**        \\
                       & (0.041)  & (0.040)     & (0.030)         \\
swear                  & -0.042   & -0.070      & -0.028          \\
                       & (0.157)  & (0.152)     & (0.115)         \\
norm\_followers\_count & -0.003   & -0.028***   & -0.024***       \\
                       & (0.008)  & (0.007)     & (0.006)         \\
masking\_True          & 0.004    & 0.005       & 0.001           \\
                       & (0.003)  & (0.003)     & (0.002)         \\
education\_True        & -0.001   & -0.001      & 0.000           \\
                       & (0.003)  & (0.003)     & (0.002)         \\
lockdowns\_True        & -0.003   & -0.003      & -0.000          \\
                       & (0.005)  & (0.005)     & (0.003)         \\
{\bf vaccines\_True}         & -0.004** & -0.005***   & -0.001          \\
                       & (0.002)  & (0.002)     & (0.001)         \\
R-squared              & 0.002    & 0.014       & 0.015           \\
R-squared Adj.         & 0.000    & 0.013       & 0.014           \\

\bottomrule
\end{tabular}
\end{center}
\caption{Full LIWC regression results of LIWC categories' impact on different types of engagement rates. Standard errors in parentheses. * $p<.1$, ** $p<.05$, ***$p<.01$}
\label{tab:liwc_regression_full}
\end{table}

\paragraph{Results for a Mixed-effect Model} The results for the mixed-effect model are shown in Table \ref{tab:mixed_effect_full}. We also attempted to fit a model with random slopes. However, the model failed to converge despite multiple attempts using different optimizers.

\begin{table}
\scriptsize
\centering
\begin{tabular}{lrrr}
\toprule
                         & \textit{Diff}    & \textit{gcl} & \textit{non-gcl}  \\
\midrule
Intercept              & 0.001    & 0.068***     & 0.068***         \\
                       & (0.001)  & (0.006)      & (0.006)          \\
i                      & 0.103*** & 0.313***     & 0.201***         \\
                       & (0.038)  & (0.036)      & (0.026)          \\
we                     & -0.014   & -0.017       & -0.007           \\
                       & (0.030)  & (0.028)      & (0.020)          \\
you                    & -0.022   & 0.010        & 0.028            \\
                       & (0.039)  & (0.037)      & (0.027)          \\
posemo                 & 0.038    & 0.037*       & 0.004            \\
                       & (0.024)  & (0.022)      & (0.016)          \\
negemo                 & -0.070   & -0.011       & 0.053*           \\
                       & (0.046)  & (0.042)      & (0.030)          \\
anx                    & 0.128*   & -0.038       & -0.152***        \\
                       & (0.069)  & (0.063)      & (0.046)          \\
anger                  & 0.106    & 0.049        & -0.047           \\
                       & (0.073)  & (0.067)      & (0.048)          \\
sad                    & 0.057    & -0.017       & -0.072           \\
                       & (0.074)  & (0.067)      & (0.049)          \\
family                 & 0.105    & 0.202**      & 0.093            \\
                       & (0.086)  & (0.079)      & (0.057)          \\
friend                 & -0.022   & 0.022        & 0.025            \\
                       & (0.093)  & (0.085)      & (0.062)          \\
body                   & -0.066   & -0.042       & 0.027            \\
                       & (0.055)  & (0.050)      & (0.036)          \\
health                 & -0.017   & -0.024       & 0.001            \\
                       & (0.026)  & (0.024)      & (0.018)          \\
work                   & 0.007    & -0.005       & -0.010           \\
                       & (0.018)  & (0.017)      & (0.012)          \\
leisure                & -0.059   & -0.043       & 0.021            \\
                       & (0.039)  & (0.036)      & (0.026)          \\
home                   & 0.019    & -0.014       & -0.031           \\
                       & (0.060)  & (0.055)      & (0.039)          \\
money                  & -0.035   & -0.045       & -0.014           \\
                       & (0.042)  & (0.039)      & (0.028)          \\
relig                  & -0.008   & -0.053       & -0.057           \\
                       & (0.081)  & (0.074)      & (0.054)          \\
death                  & 0.018    & 0.010        & -0.016           \\
                       & (0.041)  & (0.038)      & (0.028)          \\
swear                  & -0.042   & -0.106       & -0.082           \\
                       & (0.157)  & (0.144)      & (0.104)          \\
norm\_followers\_count & -0.003   & -0.145**     & -0.146*          \\
                       & (0.008)  & (0.070)      & (0.078)          \\
masking\_True          & 0.004    & 0.005**      & 0.001            \\
                       & (0.003)  & (0.003)      & (0.002)          \\
education\_True        & -0.001   & -0.002       & -0.001           \\
                       & (0.003)  & (0.003)      & (0.002)          \\
lockdowns\_True        & -0.003   & 0.001        & 0.004            \\
                       & (0.005)  & (0.004)      & (0.003)          \\
vaccines\_True         & -0.004** & -0.005**     & -0.000           \\
                       & (0.002)  & (0.002)      & (0.001)          \\
Group Var              & 0.000    & 1.169***     & 2.803***         \\
                       & (nan)    & (0.140)      & (0.302)          \\
\bottomrule
\end{tabular}
\caption{Mixed-effect regression results of LIWC categories' impact on different types of engagement rates. The group variable is the authors. Standard errors in parentheses. * $p<.1$, ** $p<.05$, ***$p<.01$}
\label{tab:mixed_effect_full}
\end{table}

\section{LLM-based Content Analysis}
\label{app:hypothesis}

We adapted the D5 framework proposed by \citet{zhong2022describing,zhong2023goal}, using the provided implementation.\footnote{https://github.com/ruiqi-zhong/D5} In our setup, we used \texttt{GPT4o-2024-08-06} as the proposer to generate hypotheses, and a provided 3B parameter version of the T5 model, as the validator to assess each hypothesis.

For each comparison, we randomly sampled 1000 tweets from each corpus to serve as the exploration (training) set used by the proposer, and another 1000 tweets per corpus as the validation set used by the validator to assign a validation score, denoted as $V'$. 
During the generation phase, the proposer randomly selects 25 tweets from each corpus in each round and continues this process until $k=20$ hypotheses are produced.

We applied this method across multiple experimental setups, varying:
(1) The percentile thresholds used to define the corpora: $20\%$ and $10\%$;
(2) The engagement metrics used for partitioning: \textit{Diff}, \textit{gcl}, and \textit{non-gcl};
(3) The level of contextual information included in the prompts.
% \begin{itemize}
% \item The percentile thresholds used to define the corpora: $20\%$ (Table~\ref{tab:diff_cf_p20}) and $10\%$ (Table~\ref{tab:diff_cf_p10});
% \item The engagement metrics used for partitioning: \textit{Diff} (Table~\ref{tab:diff_cf_p20}), \textit{gcl} (Table~\ref{tab:gcl_cf_p20}), and \textit{non-gcl} ( Table~\ref{tab:non_gcl_cf_p20});
% \item The level of contextual information included in the prompts.
% \end{itemize}

By default, we used a minimal-context prompt shown in Table~\ref{tab:prompt_f}, which only describes the task and not the source or nature of the tweets. 
The original D5 setup uses a more detailed prompt (Table~\ref{tab:prompt_d5}) that includes explicit information about the corpora. However, we observed that when context about author or participant location is included—especially without being visible in the tweet text—the LLM tends to hallucinate references to ``locality'', as seen in the generated hypotheses from Table~\ref{tab:diff_p20}.

We also tested a prompt that encourages the model to focus on linguistic and psychological patterns, broadly aligned with LIWC categories (prompt shown in Table~\ref{tab:prompt_new_target}
; results in Table~\ref{tab:diff_cf_p20_new_target}). While this prompt adjustment resulted in slightly different hypotheses, the overall content and distributional patterns remained similar to the default prompt (Table~\ref{tab:diff_cf_p20}).

Similarly, we found only minor differences between results generated using 20\% vs. 10\% percentile thresholds.
(Table~\ref{tab:diff_cf_p20} and Table~\ref{tab:diff_cf_p10})
Across both setups, dominant hypotheses continued to emphasize the sharing of personal experience and emotional expression.
Finally, the comparison between \textit{gcl}-based and \textit{non-gcl}-based groupings 
(Tables~\ref{tab:gcl_cf_p20} and \ref{tab:non_gcl_cf_p20})
yields consistent themes—namely, the salience of emotional expression and first-person narratives—echoing the patterns observed in our regression analysis (Table~\ref{tab:liwc_regression}).
These features are significant predictors across both \textit{gcl} and \textit{non-gcl} setups, and remain robust even when analyzing the \textit{Diff} metric, which controls for shared effects across both dimensions.

Overall, the LLM-based content analysis complements our statistical findings by surfacing interpretable hypotheses that support the observed role of personal expression and affective language in shaping engagement.

\begin{table}[htbp]
\centering
\small
\begin{tcolorbox}[title={Prompt used for hypothesis generation without \textit{gcl} context}, colback=gray!5, colframe=gray!50!black, fonttitle=\bfseries]
Group A: \{sampled tweets from Corpus A\} \\
Group B: \{sampled tweets from Corpus B\}\\

The dataset includes two sets of tweets with different engagement metrics. The two groups are generated based on different tweet engagement metrics. Group A contains tweets in Group A, while Group B contains tweets in Group B.\\

I am a public health communication researcher. My goal is to figure out what kind of tweets are more frequent in Group A compared to Group B.\\

Please write a list of hypotheses about the datapoints from Group A (listed using bullet points "-"). Each hypothesis should be formatted as a sentence fragment.

Based on the two sentence groups (A and B) above, more sentences in Group A ...
\end{tcolorbox}
\caption{Prompt used for hypothesis generation}
\label{tab:prompt_f}
\end{table}

\begin{table}[htbp]
\centering
\small
\begin{tcolorbox}[title={Prompt used for hypothesis generation with \textit{gcl} context}, colback=gray!5, colframe=gray!50!black, fonttitle=\bfseries]
Group A: \{sampled tweets from Corpus A\} \\
Group B: \{sampled tweets from Corpus B\}\\

The dataset includes tweets with various geo-co-located and non-geo-co-located engagement rates. The two groups are generated based on the difference between geo-co-located and non-geo-co-located engagement. Group A contains tweets with higher geo-co-located engagement rate but lower non-geo-co-located engagement rate, while Group B contains tweets with higher non-geo-co-located engagement rate but lower geo-co-located engagement rate.\\

I am a public health communication researcher. My goal is to figure out what kind of tweets are more frequent in tweets with higher geo-co-located engagement rate but lower non-geo-co-located engagement rate.\\

Please write a list of hypotheses about the datapoints from Group A (listed using bullet points "-"). Each hypothesis should be formatted as a sentence fragment.

Based on the two sentence groups (A and B) above, more sentences in Group A ...
\end{tcolorbox}
\caption{Prompt used for hypothesis generation {\bf with} \textit{gcl} context}
\label{tab:prompt_d5}
\end{table}

\begin{table}[htbp]
\centering
\small
\begin{tcolorbox}[title={Prompt encouraging hypotheses aligned with LIWC categories}, colback=gray!5, colframe=gray!50!black, fonttitle=\bfseries]
Group A: \{sampled tweets from Corpus A\} \\
Group B: \{sampled tweets from Corpus B\}\\

The dataset includes two sets of tweets with different engagement metrics. The two groups are generated based on different tweet engagement metrics. The Group A snippets tweets in Group A, while the Group B snippets tweets in Group B.  \\

I am a public health communication researcher. My goal is to figure out what features related to {\bf linguistic patterns, topics of discussion, and psychological processes} are more frequent in Group A compared to Group B.\\

Please write a list of hypotheses about the datapoints from Group A (listed using bullet points "-"). Each hypothesis should be formatted as a sentence fragment.

Based on the two sentence groups (A and B) above, more sentences in Group A ...
\end{tcolorbox}
\caption{Prompt encouraging hypotheses aligned with LIWC categories}
\label{tab:prompt_new_target}
\end{table}

\begin{table}[htbp]
\centering
\begin{tcolorbox}[title={Prompt for profession classification task}, colback=blue!5, colframe=blue!50!black, fonttitle=\bfseries]
For the following Twitter user profile description and username, infer whether the user's profession is related to each of the professions as follows: media professions (e.g., journalist, reporter, writer, editor, and author), academic (e.g., professor, researcher, and scientist), health professions (e.g., physician, doctor, epidemiologist), and politicians. For each Twitter user, output the results for each of the four professions and provide a brief explanation in the JSON format of: 

\texttt{\{"screen\_name": USER\_NAME, "Media\_Professions": true / false, "Academic": true / false, "Health\_Professions": true / false, "Politicians": true / false, "Explanation": "[Explanation]"\}} \\

Only output \texttt{true} for the best-matched professions if possible. All outputs should be in the JSON format as described above.

username: \{screen\_name\} \\
description: \{description\}

\end{tcolorbox}
\caption{Prompt used for classifying professions of Twitter users}
\label{tab:prompt_profession}
\end{table}

{
\renewcommand{\arraystretch}{1.3} % adds vertical padding between rows
\begin{table}[ht]
\centering
\small
\begin{tabularx}{\linewidth}{>{
\arraybackslash}X r}
\toprule
\textbf{Hypothesis} & $V'$ \\
\midrule
include expressions of personal emotions or sentiments, such as fear, pride, or frustration & 0.081 \\
focus on personal experiences and anecdotes related to the pandemic & 0.070 \\
- discuss personal anecdotes and experiences related to COVID-19 & 0.063 \\
discuss everyday life adjustments and challenges during the pandemic, like dating or travel & 0.061 \\
incorporate personal milestones or special occasions and how they've been affected by the pandemic & 0.055 \\
reference specific locations or cities, such as Scottsdale, AZ or Ohio & 0.043 \\
mention local government actions and leadership, especially in specific states or cities & 0.039 \\
describe impacts of COVID-19 measures on daily life and personal routines & 0.037 \\
contain direct calls to action or emphasize the importance of community actions (e.g., wearing masks or supporting essential workers) & 0.024 \\
address COVID-19 misinformation and actions taken by social media platforms & 0.020 \\
mention specific public health measures like mask mandates & 0.015 \\
focus on COVID-19 vaccine distribution and individual vaccination experiences & 0.008 \\
talk about the reopening or closure of specific facilities, such as nursing homes & 0.005 \\
highlight stories of specific individuals or public figures contracting COVID-19 & 0.000 \\
report specific COVID-19 statistics or data related to local settings like neighborhoods or cities & -0.004 \\
discuss innovations or changes in healthcare or public health practices due to COVID-19 & -0.008 \\
highlight the impact of COVID-19 on individual or specific communities (e.g., hospitals being full, grocery store experiences) & -0.025 \\
mention specific public figures or local leaders in the context of COVID-19 response & -0.032 \\
mention trends or statistics related to COVID-19 cases or immunity & -0.061 \\
include updates or opinions about scientific research and data & -0.120 \\
\bottomrule
\end{tabularx}
\caption{LLM-generated distributional difference without \textit{gcl} context: tweets bottom 20\% percentile of the \textit{Diff}values vs. the top 20\% percentile of tweets. Full results for Table \ref{tab:diff_cf_p20_short}.}
\label{tab:diff_cf_p20}
\end{table}

\begin{table}[ht]
\centering
\small
\begin{tabularx}{\linewidth}{>{
\arraybackslash}X r}
\toprule
\textbf{Hypothesis} & $V'$ \\
\midrule
include humorous or light-hearted content related to coping with the pandemic & 0.147 \\
incorporate humor or light-hearted personal commentary about adjusting to pandemic life & 0.146 \\
- express personal experiences or anecdotes related to the COVID-19 pandemic & 0.135 \\
include casual language, emojis, or express personal emotions & 0.097 \\
reflect on the emotional or psychological impacts of COVID-19 & 0.071 \\
...... &  \\
mention specific locations or regions affected by the pandemic & -0.054 \\
\bottomrule
\end{tabularx}
\caption{LLM-generated distributional difference without \textit{gcl} context: tweets
with 0\% \textit{gcl} engagement rate vs. the top 20\% percentile of tweets. }
\label{tab:gcl_cf_p20}
\end{table}

\begin{table}[ht]
\centering
\small
\begin{tabularx}{\linewidth}{>{
\arraybackslash}X r}
\toprule
\textbf{Hypothesis} & $V'$ \\
\midrule
emphasize personal or anecdotal perspectives on pandemic-related issues & 0.164 \\
- focus on personal opinions and reflections related to COVID-19 experiences & 0.128 \\
express personal frustrations or emotions related to COVID-19, including feelings of worry or anger & 0.123 \\
engage in casual or conversational tone, often posing questions to the audience & 0.092 \\
include direct responses or criticisms toward specific political figures, particularly Donald Trump & 0.070 \\
...... & \\
draw comparisons with other countries or states to emphasize pandemic management or outcomes & -0.068 \\
reflect the state-level impact of COVID-19, such as hospital capacities in Wisconsin or actions by California & -0.093 \\
share insights on localized public health measures, like vaccination appointments in specific areas & -0.103 \\
include hyperlinks or references to full articles or studies, possibly indicating a focus on data and evidence & -0.114 \\
\bottomrule
\end{tabularx}
\caption{LLM-generated distributional difference without \textit{gcl} context: tweets
with 0\% \textit{non-gcl} engagement rate vs. the top 20\% percentile of tweets.}
\label{tab:non_gcl_cf_p20}
\end{table}

\begin{table}[ht]
\centering
\small
\begin{tabularx}{\linewidth}{>{
\arraybackslash}X r}
\toprule
\textbf{Hypothesis} & $V'$ \\
\midrule
feature personal stories and specific events & 0.085 \\
cover anecdotes about personal interactions or encounters related to COVID-19 & 0.073 \\
discuss the implications of COVID-19 on {\bf local healthcare systems or hospitals} & 0.043 \\
report on {\bf local vaccination efforts or challenges} & 0.036 \\
...... & \\
focus on domestic political issues and figures, such as the Trump administration and state-level responses & -0.050 \\
focus on regional COVID-19 statistics and updates & -0.052 \\
\bottomrule
\end{tabularx}
\caption{LLM-generated distributional differences {\bf with} \textit{gcl} context: tweets with bottom 20\% percentile \textit{Diff} values vs. top 20\% percentile of tweets. Hallucinations about "locality" are bolded}
\label{tab:diff_p20}
\end{table}

\begin{table}[ht]
\centering
\small
\begin{tabularx}{\linewidth}{>{
\arraybackslash}X r}
\toprule
\textbf{Hypothesis} & $V'$ \\
\midrule
share light-hearted or personal sentiments about adapting to the pandemic & 0.072 \\
use humor or cultural references in relation to the pandemic & 0.063 \\
...... & \\
address misinformation or controversies related to COVID-19 testing, vaccines, or treatments & -0.052 \\
report on statistics, forecasts, or models related to COVID-19 cases and mortality rates & -0.065 \\
mention high-profile public figures in the context of pandemic-related policies & -0.094 \\
focus on COVID-19 vaccination efforts or challenges & -0.112 \\
discuss the ongoing effects and responses to COVID-19 in specific geographic regions or countries & -0.140 \\
\bottomrule
\end{tabularx}
\caption{LLM-generated distributional differences without \textit{gcl} context: tweets with bottom 10\% percentile \textit{Diff} values vs. top 10\% percentile of tweets.}
\label{tab:diff_cf_p10}
\end{table}

\begin{table}[ht]
\centering
\small
\begin{tabularx}{\linewidth}{>{
\arraybackslash}X r}
\toprule
\textbf{Hypothesis} & $V'$ \\
\midrule
use a conversational or colloquial tone & 0.082 \\
capture personal or anecdotal experiences & 0.075 \\
...... & \\
\bottomrule
\end{tabularx}
\caption{LLM-generated distributional difference without \textit{gcl} context, focusing on linguistic patterns, topics, and psychological processes: tweets with bottom 20\% percentile \textit{Diff} values vs. top 20\% percentile of tweets.}
\label{tab:diff_cf_p20_new_target}
\end{table}

}

\section{Supplement Results for Section \ref{sec:reply}}
\label{app:reply}

\begin{figure*}[!htb]
    \centering
    \includegraphics[scale=0.29]{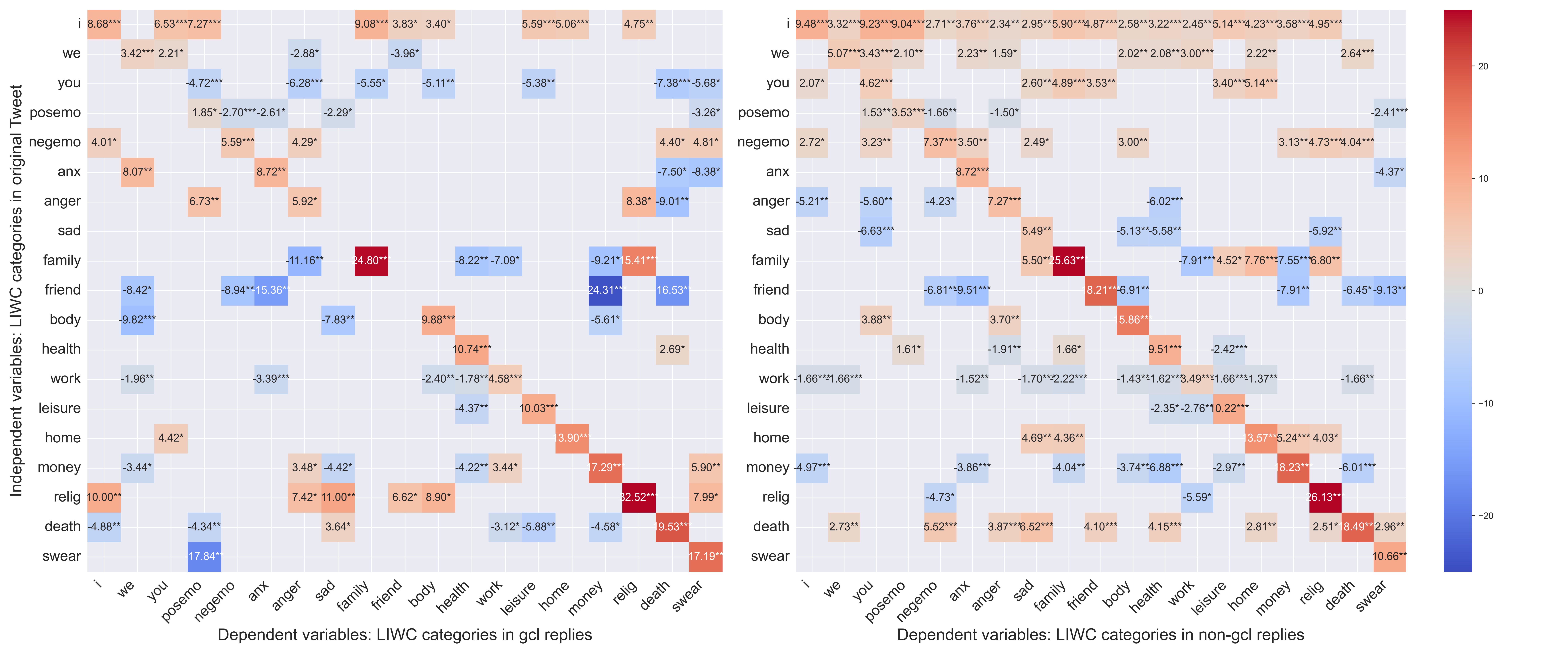}
    \caption{How \textit{gcl} replies and \textit{non-gcl} with specific LIWC categories correlate with LIWC features in the original Tweet. Each column is a logistic regression. Each cell is the coefficient. $p < 0.1$ is marked with *, $p < 0.05$ with **, and $p < 0.01$ with ***. }
    \label{fig:heatmap_full}
\end{figure*}

Table \ref{tab:logit} shows the logistic regression results for comparing \textit{gcl} vs. \textit{non-gcl} replies. 
% Table \ref{tab:gcl_replies} shows the hypotheses generated by the LLM-based method.
The heatmaps with all LIWC categories from Section \ref{sec:reply} are shown in Figure \ref{fig:heatmap_full}.
{
\renewcommand{\arraystretch}{1.3}

\begin{table}[tb]
\centering
\small
\begin{tabularx}{\linewidth}{>{\arraybackslash}X r}
\toprule
\textbf{Hypothesis} & \textbf{$V'$} \\
\midrule
convey frustration with the impact of legislation or policies on workers or specific groups & 0.040 \\
include a call to action or advocacy, such as petitions or protests & 0.029 \\
critique or express dissatisfaction with the federal government's response to the pandemic & 0.027 \\
express support or admiration for local leaders or officials & 0.014 \\
reference specific political events, policies, or government figures, often critically & 0.013 \\
discuss American politics or criticize political figures & 0.011 \\
mention COVID-19 in the context of state or federal government response or criticism & 0.010 \\
engage in direct confrontation or argument with other users & 0.010 \\
express gratitude or appreciation for someone's efforts or leadership & 0.009 \\
express frustration or emotional reactions, often in a casual or informal tone & 0.008 \\
reference data or statistics related to COVID-19 cases and mortality rates & 0.008 \\
debate the effectiveness of public health guidelines or actions & 0.007 \\
inquire about or critique the implementation of healthcare measures or policies & 0.005 \\
exhibit colloquial language or informal expressions & 0.005 \\
discuss the importance or implications of COVID-19 precautions, such as mask-wearing & 0.001 \\
mention political figures in the context of their response to the pandemic & -0.001 \\
reflect concern for the long-term effects of COVID-19 or other health-related issues & -0.001 \\
use sarcasm or mocking language & -0.005 \\
contain personal insults or hostile language & -0.007 \\
\bottomrule
\end{tabularx}
\caption{LLM-generated distributional difference without \textit{gcl} context: \textit{gcl} replies vs. \textit{non-gcl} replies.}
\label{tab:gcl_replies}
\end{table}
}

\section{Prompt for Profession Classification}
\label{app:prompt}

The prompt we used for profession classification in Section~\ref{sec:user} is shown in Table~\ref{tab:prompt_profession}. We instructed the LLM to infer whether each user's profile suggested affiliation with one or more of four target professions: media, academic, health, or political. The model was asked to output a binary judgment for each profession, along with a brief explanation, using a structured JSON format.

Note that \texttt{\{screen\_name\}} and \texttt{\{description\}} are replaced with actual usernames and descriptions. A sample output from ChatGPT with an anonymized user name is:\\
\texttt{
\{"screen\_name": "ANONYMIZED", "Media\_Professions": false, "Academic": true, "Health\_Professions": true, "Politicians": false, "Explanation": "The user is a Clinical Psychology PhD Student and does research in Global Mental Health, which indicates that he is involved in academic and health professions."\}
}

\end{document}